\documentclass[aps,prb,superscriptaddress,twocolumn,floatfix]{revtex4}

\usepackage[utf8]{inputenc}
\usepackage{mathtools}
\usepackage{amsmath, amsthm, amssymb, amsfonts}
\usepackage{hyperref}
\usepackage{graphicx}
\usepackage{xcolor}
\usepackage{comment}

\hypersetup{
  colorlinks=true,
  citecolor=blue,
  linkcolor=blue,
  urlcolor=blue,
  }

\begin{document}


\title{Exploring the magnetic states in the one-band Hubbard model:\\
  Impact of long-range hoppings}

\author{Sudip Mandal}
\affiliation{Theory Division, Saha Institute of Nuclear Physics,
  A CI of Homi Bhabha National Institute, Kolkata 700064, India}

\author{Sandip Halder}
\affiliation{Theory Division, Saha Institute of Nuclear Physics,
  A CI of Homi Bhabha National Institute, Kolkata 700064, India}

\author{Kalpataru Pradhan}
\email{kalpataru.pradhan@saha.ac.in}
\affiliation{Theory Division, Saha Institute of Nuclear Physics,
  A CI of Homi Bhabha National Institute, Kolkata 700064, India}

\date{\today}

\begin{abstract}
Correlated electron systems with competing interactions provide a valuable
platform for examining exotic magnetic phases. Theoretical models often focus on
nearest-neighbor interactions, although long-range interactions can have a considerable
impact on the behavior of the system, creating distinct and complicated magnetic phases.
We investigate the consequences of competing interactions in a half-filled one-band
Hubbard model on a simple cubic lattice, incorporating hopping processes up to
third-nearest-neighbors, to explore the underlying magnetotransport properties. Our
magnetic phase diagrams at low temperatures, obtained using semi-classical Monte Carlo
analysis, reveal that the long-range interactions can disrupt one form of magnetic
phase while creating a new type of magnetic order. For the nonperturbative regime
(on-site Hubbard repulsive strength $U \sim$ bandwidth) the C-type antiferromagnetic ground
state is preferred over the G-type antiferromagnetic phase when the interaction
between second-nearest neighbor sites becomes significant to the nearest-neighbor 
interactions. However, interactions beyond the second-nearest-neighbors are required to
stabilize the A-type antiferromagnetic ground state. Remarkably, at low temperatures,
a highly correlated paramagnetic insulating phase develops at the intersection
between the antiferromagnetic phases, which might promote a three-dimensional
spin-liquid state.
\end{abstract}

\maketitle

\section{Introduction}
Strong electron-electron interactions, a defining feature of correlated electron
systems, result in emergent behaviors~\cite{Lacroix, Dagotto, Sachdev} that deviate
widely from those predicted by traditional band theory~\cite{Herman}. These quantum
materials exhibit remarkable phenomena~\cite{Lacroix, Sachdev} including unconventional
superconductivity~\cite{Anderson, Norman, Liu, Keimer}, huge magnetoresistance~\cite{Rao, Buschow},
and novel magnetic orderings~\cite{Fazekas, Sachdev2, Balents, Lee}. In order to
comprehend the intricate properties of highly correlated material systems, one needs
to analyze complex mechanisms underlying the interplay of strong electron-electron
interactions, spin, charge, orbital degrees of freedom, and the underlying lattice
structure~\cite{Dagotto, Witczak, Soumyanarayanan}. The understanding of these systems
has been enhanced by recent advances in theoretical frameworks such as dynamical mean-field
theory (DMFT)~\cite{Georges, Rohringer, Sato, Fratino},
quantum Monte Carlo (QMC)~\cite{Rohringer, Blankenbecler, Staudt, Foulkes, Senechal}, and
tensor network approaches~\cite{Corboz, Banuls, Vlaar, Vlaar2}, as well as 
experimental techniques like high-resolution spectroscopy~\cite{Cavalieri, Valmispild}, 
scanning tunneling microscopy~\cite{Wiesendanger, Eickhoff}, 
and ultracold-atom simulations~\cite{Gross, Schafer, Langen}.

Investigating the magnetic ordering is essential for identifying a wide range
of exotic phenomena in quantum materials~\cite{Lacroix, Goyal}. In fact, the onset and
suppression of the long-range antiferromagnetic (AF) order are the most studied
phenomena in correlated systems~\cite{Staudt, Raczkowski, Langmann, Laubach, Fratino2, Jana, Chakraborty}.
Specifically, AF order has been thoroughly examined in materials with rich phase
diagrams~\cite{Staudt, Langmann, Chakraborty} and unusual characteristics~\cite{Laubach, Jana},
such as cuprates~\cite{Liu2, Delannoy}, iron-pnictides~\cite{Krellner, Delacruz}, 
and iron-chalcogenides~\cite{Liu}. Recently, ultracold-atom experiments have become
increasingly effective methods for examining AF correlations in controlled
settings~\cite{Jaksch, Hart, Mazurenko}. These systems enable accurate adjustment
of interaction strengths and geometries by simulating the behavior of strongly
correlated electrons in optical lattices~\cite{Gross, Schafer, Langen, Greiner}.
The identification of antiferromagnetic correlations at low-temperatures in
repulsively-interacting Fermi gases~\cite{Sanner, Ji} is the significant one among the notable accomplishments.

Ongoing research focuses on the impact of competing interactions on the magnetic
characteristics and phase diagrams of pristine systems~\cite{Laubach, Jana, Kim, fujihala, lee3}.
These conflicting interactions have the potential to disturb long-range magnetic order
and produce new intricate magnetic phases~\cite{Laubach, Jana, fujihala}. For example,
competitive interactions can arise from geometrically incompatible spin alignments, such
as those found in triangular~\cite{Fazekas, Moessner, Itou, Arh, Bu, Bairwa}
or Kagome lattices~\cite{Lee2, Helton, Mendels}, which may hinder the formation of
traditional long-range antiferromagnetic order and instead result in exotic states
like non-collinear spin arrangements or spin liquids. Long-range interactions in
non-frustrated systems can lead to complicated phase boundaries with competing
magnetic ordering, creating intricate phase diagrams~\cite{Raczkowski, Laubach, Jana, Becca}.
Actually, second-nearest-neighbor interactions have a major impact on how the
charge, spin, and orbital degrees of freedom interact in transition metal-based oxide
materials~\cite{Raczkowski, Delannoy, Shenoy, Rak}.

As a fundamental framework for studying and assessing various approaches that
characterize different kinds of correlated electron systems, the Hubbard model 
continues to be of interest~\cite{Gutzwiller, Kanamori, Hubbard}.  
It is particularly helpful for understanding phenomena like magnetism and 
metal-insulator transitions~\cite{Staudt, Chakraborty, Kondo, Igoshev}. 
Therefore, despite its seemingly simple specification, the model provides a
versatile platform for examining the emergence of intricate collective phenomena
that are influenced by temperature, band structure, interaction strength, and
electron filling~\cite{Hirsch, Laubach, Jana, Chakraborty}. These phenomena
are caused by the complex interaction of
correlations and ordering tendencies, which can compete or cooperate in
fascinating ways. While nearest-neighbor interactions are frequently highlighted
in theoretical approaches, the addition of second and third nearest-neighbor
interactions can have a significant impact on the dynamics of the system,
adding new phases and enhancing the complexity of its phase
diagrams~\cite{Langmann, Laubach, Becca, Lin}. In fact, long-range hopping
processes in the Hubbard model may provide a more thorough and potentially realistic
description of some specific materials.

The ground state features are changed when electrons are allowed to hop to 
second-nearest-neighbor sites because this creates competition between nearest and 
second-nearest-neighbor hoppings~\cite{Jana, Langmann, Lin, Laubach}. For instance, in
relation to the half-filled Hubbard model, nearest-neighbor interactions typically
favor G-type AF ordering whereas the second-nearest-neighbor interactions disrupt
G-type antiferromagnetic order, causing spin rearrangement. As a result, the
C-type antiferromagnetic arrangement may be preferred over the G-type
antiferromagnetic state when the interaction between the second-nearest-neighbor
spins becomes significant. At the same time, interactions beyond the
second-nearest-neighbors are necessary to stabilize the A-type antiferromagnetic
ground state~\cite{Laubach}. The strength of the long-range hopping parameters,
therefore, determines the particular kind of magnetic order that is seen as the
ground state. However, these conflicting interactions can also create a situation
where exchange interactions can not be satisfied at the same time, resulting in
the emergence of a correlated paramagnetic insulating state~\cite{Laubach, Becca, Tocchio}
at low temperatures. The combination of paramagnetic insulating state and long-range
interactions creates conditions that are favorable for the creation of spin
liquid states~\cite{Laubach, Becca}.

Although long-range interactions typically weaken with distance, they nevertheless have
a significant impact on the properties of the system. However, the understanding
of the magnetotransport features of the half-filled Hubbard model with
long-range interactions in three dimensions remains limited. In this work, we examine
how crucial second- and third-nearest-neighbor interactions are in promoting magnetic
transitions in the context of highly coupled electron systems. On a simple cubic lattice,
we concentrate on the $t$-$t'$-$t''$ based Hubbard model, a paradigmatic approach
that includes hopping amplitudes for generating nearest ($t$), second-nearest ($t'$),
and third-nearest-neighbors ($t''$) interactions. We investigate the competition
among $t$, $t'$, and $t''$ that causes frustration, modifies band structures, and
affects the magnetotransport characteristics of the systems by systematically
adjusting the respective strengths of these hopping parameters. In order to examine
the ground states and their changes in relation to temperature ($T$) and Hubbard
interaction strength ($U$), we use a well-developed semi-classical Monte Carlo (s-MC)
method~\cite{Jana, Chakraborty, Mukherjee, Mukherjee2, Tiwari}. We mainly focus on
the nonperturbative regime (U $\sim$ bandwidth) in our study. The nearest-neighbor
interaction establishes the G-type AF phase, while the second-nearest-neighbor
interaction disrupts it and sets up the C-type AF phase although the insulating nature
remains intact. In addition, our calculations show that the combined effect of
three types of hoppings ($t$, $t'$, and $t''$) stabilizes the A-type AF ground
state. Remarkably, a small but finite paramagnetic insulating region emerges from the
combined action of the first-, second-, and third-nearest-neighbor interactions
at the junction of the three types of magnetic ground state phases which can
potentially host a spin-liquid phase.

This paper is organized as follows: In Sec.~\textbf{II}, we outline the model
Hamiltonian, long-range hopping parameters, and numerical technique.
Appendix~\ref{derivation_Heff} provides more information on the model Hamiltonian.
In Appendix~\ref{obs}, we outline the calculation of various physical observables
used to investigate magnetotransport properties. At first, we briefly discuss
about the magnetotransport properties of the reference Hubbard model in
Sec.~\textbf{III}. Next, we focus on how the
second-nearest interaction affects the system's magnetotransport features in
Sec.~\textbf{IV}. The combined effect of the second- and third-nearest-neighbor
interactions is then investigated in Sec.~\textbf{V}. At the end, we
present a summary of our results in Sec.~\textbf{VI}.

\section{Model Hamiltonian and Method}
We explore the ground state magnetic phases for the following
one-band Hubbard Hamiltonian:

\begin{align} \label{hamiltonian}
H  =& -t \!\! \sum_{\left\langle i,j \right\rangle , \sigma} \!\! c_{i,\sigma}^\dagger c_{j,\sigma} -t' \!\!\!\! \sum_{\left\langle \left\langle i,j \right\rangle \right\rangle, \sigma} \!\!\!\! c_{i,\sigma}^\dagger c_{j,\sigma}-t'' \!\!\!\! \sum_{\left\langle \left\langle \left\langle i,j \right\rangle \right\rangle \right\rangle, \sigma} \!\!\!\! c_{i,\sigma}^\dagger c_{j,\sigma} \nonumber \\
& + U \sum_i n_{i\uparrow} n_{i \downarrow} - \mu \sum_i n_i \\
=& H_0 +H_I \nonumber
\end{align}

where $t$, $t'$, and $t''$ are the nearest-neighbor, second-nearest-neighbor,
and third-nearest-neighbor hopping  parameters, respectively.
$\langle\rangle$, $\langle\langle\rangle\rangle$, and $\langle\langle\langle\rangle\rangle\rangle$
correspond to the first-, second-, and third-nearest-neighbor sites, respectively.
$c_{i,\sigma}$ ($c_{i,\sigma}^\dagger$) represents the fermion annihilation
(creation) operator with spin $\sigma$ at site $i$.
$n_{i}$ ($ = \sum_{\sigma}c_{i,\sigma}^\dag c_{i,\sigma}$) is the number operator.
$U$ ($ > 0$) represents the strength of on-site repulsive Hubbard interaction.
$\mu$ is the chemical potential, which controls the overall average density of the system.
For $t' = 0$, the spectrum is particle-hole symmetric and $\mu$ turns out to be zero
for half-filled case. However, in the same half-filled case, $\mu$
becomes finite for nonzero $t'$ and $t''$. In our definitions, $H_{0}$
represents the non-interacting quadratic part of the model Hamiltonian, whereas
$H_{I}$ containing the interacting quartic term. Our model Hamiltonian calculations
are based on the simple cubic lattice that includes periodic boundary conditions.

To solve the model Hamiltonian, we decompose the interacting quartic term ($H_{I}$)
of the model Hamiltonian by implementing the standard Hubbard-Stratonovich transformation
by introducing Hubbard-Stratonovich (HS) auxiliary fields ($\mathbf{m}_i$ and $\phi_{i}$)
at an arbitrary site $i$.  This allows us to employ the semi-classical Monte Carlo (s-MC) 
approach to simulate the Hamiltonian. In this decomposition, the vector auxiliary field
($\mathbf{m}_i$) couples with the spin degrees of freedom, while the  scalar auxiliary
field ($\phi_i$) couples with the charge degrees of freedom. Next, we consider the
auxiliary fields to be time-independent and treat them as classical fields. We set
the scalar field $\phi_i$ to be proportional to the average density $<n_{i}>$ and
define  $i \phi_i = \frac{U}{2} \left\langle n_i \right\rangle$ at the saddle-point
level. However, the auxiliary fields are thermally fluctuating and spatially
inhomogeneous, which is the hallmark of our model to capture phases as finite
temperatures. Here, it is important to note that the HS transformation is a local
transformation that decouples the many-body on-site interaction. Using these 
approximations (for details, see appendix~\ref{derivation_Heff}), we obtain the
following effective spin-fermion Hamiltonian~\cite{Chakraborty, Mukherjee, Halder, Bidika}:
\begin{align}\label{h_eff}
H_{eff} = & H_0 + \frac{U}{2} \sum_i \left( \left\langle n_i \right\rangle n_i - \mathbf{m}_i.{\boldsymbol{\sigma}}_i \right) \nonumber \\ 
 & + \frac{U}{4} \sum_i \left( \mathbf{m}_i^2 - \left\langle n_i \right\rangle^2 \right) - \mu \sum_i n_i,
\end{align}
where $H_0$ represents the  hopping terms of the Hamiltonian.

We  solve the effective Hamiltonian ($H_{\text{eff}}$) using the exact diagonalization
based s-MC technique. In this approach, we diagonalize the Hamiltonian
for a fixed configuration of auxiliary fields $\{\mathbf{m}_i\}$ and average charge
densities $\{\left\langle n_i \right\rangle\}$. At a fixed temperature, the auxiliary fields
$\{\mathbf{m}_i\}$ are annealed by employing the Metropolis algorithm. For a fixed temperature,
we use 2000 Monte Carlo system sweeps for annealing: first half of the MC sweeps are used to
thermally equilibrate the system and the rest MC sweeps are used for measuring the physical
observables. During annealing, the average electron densities $\{\left\langle n_i \right\rangle\}$
are updated self-consistently at every $10th$ Monte Carlo step. The main objective of this process
is to generate an equilibrium configuration of the auxiliary fields $\mathbf{m}_i$ and the
average electron densities $\left\langle n_i \right\rangle$. $\mu$ is also evaluated
self-consistently during the Monte Carlo sweeps to ensure that the system remains at
half-filling. Then, we  calculate the expectation values of the observables using the
eigenvectors and eigenvalues, obtained from diagonalizing the Hamiltonian in equilibrium
configuration. These individual expectation values from the equilibrium configurations
are further averaged over the results from 100 such configurations at a fixed temperature. 
All observables are determined by averaging the values obtained from individual
configurations. It is important to note that we calculate observables from every $10th$ 
equilibrium configuration in order to avoid spurious self-correlations. The temperature
is gradually lowered in small steps to allow for equilibration. To overcome size
limitations, we employ the Monte Carlo technique within a traveling cluster approximation 
(TCA)~\cite{Kumar, Chakraborty, Chakraborty2, Bidika, Halder2} to handle system sizes of $L^3=10^3$, where $L$ is the length
of the sides of the simple cubic lattice. To ensure consistency, throughout our s-MC study, 
we assume $t = 1$ and $t'$, $t''$ are expressed with respect to $t$.
We also express $U$, $T$ (temperatures) in units of $t$. 

We compute a number of physical observables in order to understand the finer details
of the magnetotransport characteristics. To study magnetic features like magnetic
transition, we calculate local moments $M$ and quantum correlations using the structure
factor $S(\textbf{q})$. Additionally, we compute temperature-dependent specific heat
$C_v$ in order to correlate the low temperature peak structure observed in $C_v$ with
the magnetic transition. Transport quantities are investigated by calculating
resistivity $\rho$ [inverse of the $dc$ limit of conductivity $\sigma(\omega)$] and
density of states (DOS). The appendix~\ref{obs} provides an outline of the definitions
of the different observables.

\section{Brief description of the magnetotransport properties of the reference Hubbard model ($t'$ = 0 and $t''$ = 0)}

\begin{figure}[t]
\centering
\includegraphics[width=0.48\textwidth]{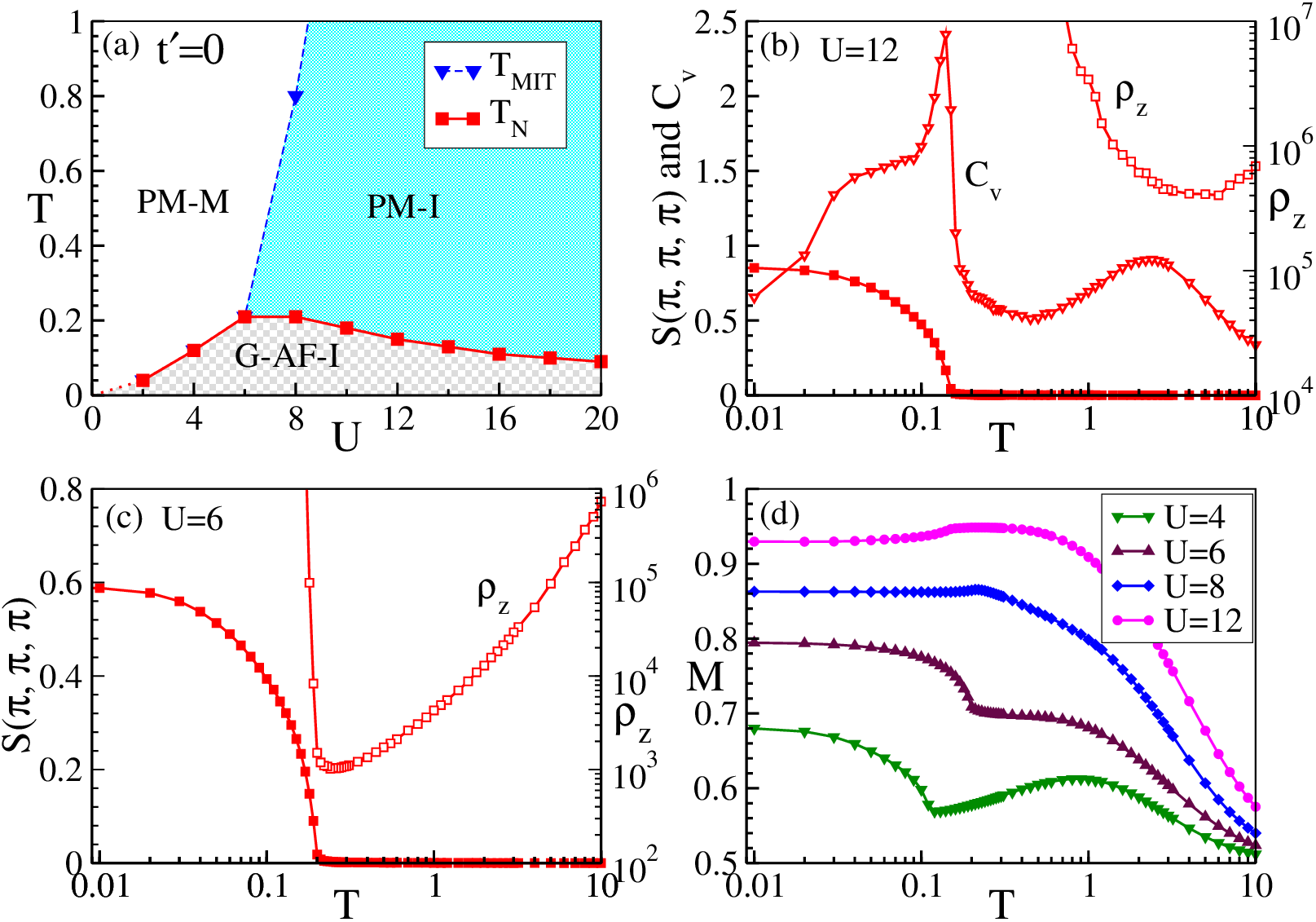}
\caption{
A brief description of the magnetotransport properties of the reference Hubbard model
($t'=0$ and $t''=0$): (a) The $U$-$T$ phase diagram: The system shifts directly from
a paramagnetic metallic (PM-M) state to a G-type AF insulating (G-AF-I) state for $U <8 $
when the temperature decreases in this regime. For $U \geq 8$, a paramagnetic insulating (PM-I)
phase exists between the PM-M and G-type AF insulating phases.
(b) Plots of $S(\pi,\pi,\pi)$ vs $T$ and $\rho_z$ vs $T$ demonstrate that for $U = 12$
($\sim$ bandwidth) the $T_{MIT}$ is greater than $T_{N}$, resulting in a PM-I region
between the G-type AF insulating and PM-M phases, as illustrated in (a). While the low
temperature peak in the specific heat ($C_v$) signifies the onset of long-range magnetic
ordering in the system, the high temperature peak, which is a marker of local moment
formation, coincides with the metal-insulator transition temperature ($T_{MIT}$).
(c) For $U = 6$, $S(\pi,\pi,\pi)$ versus $T$ and $\rho_z$ vs $T$ show that the magnetic
transition and the MIT happen at the same temperature, i.e. $T_N = T_{MIT}$.
(d) As the temperature decreases, the local magnetic moment $M$ increases. For $U \ge 8$,
$M$ saturates at low temperatures, with a minor peak around $T_N$. For lower $U$ values
$M$ initially remains relatively steady (for $U = 6$) or drops (for $U = 4$) from
$T \approx 1$ to $T \approx T_N$, before increasing below $T_N$.
}
\label{fig01}
\end{figure}

First, we briefly address the physics of the reference Hubbard model without long-range
hoppings (i.e., $t' = 0$ and $t'' = 0$) before looking into the new phenomena observed
for $t' \neq 0$ and/or $t'' \neq 0$ cases. We illustrate the $U$-$T$ phase diagram for
the reference model in Fig.~\ref{fig01}(a). At low temperatures, the ground state
displays a G-type AF insulating phase for all $U$ values similar to previous results 
in 3D~\cite{Rohringer, Staudt, Kent}. Importantly, the Neel temperature ($T_N$)
behaves non-monotonically as a function of $U$, as discussed in earlier works using
various techniques~\cite{Fratino, Staudt, Mukherjee}. The G-type AF insulating
phase persists even at $U \to 0$ limit although the resulting $T_N$ is very small.
The $T_N$ rises with $U$ until $U = 8$, then decreases as $U$ increases. For $U < 8$,
the long-range G-type AF insulating phase transits directly to the paramagnetic
metallic phase as the temperature increases. So, the metal-insulator transition
temperature ($T_{MIT}$) coincides with $T_N$. For $U \geq 8$, a paramagnetic
insulating (PM-I) phase separates the G-type AF and Paramagnetic metallic phases.
This PM-I phase is characterized by preformed local moments, 
where the thermal fluctuations suppress long-range order~\cite{Mukherjee}.

For the analysis of magnetic and transport properties, we use $U = 6$ and $U = 12$, from the
two distinct regimes in the $U$-$T$ phase diagram. $U = 6$ and $U = 12$ also correspond to
$U/\text{BW} = 0.5$ and $1$, respectively, where $\text{BW}$ denotes the non-interacting
bandwidth. Figures~\ref{fig01}(b) and (c) show the
temperature dependence of the magnetic structure factor $S(\pi, \pi, \pi)$ and resistivity
$\rho$ for $U = 12$ and $U = 6$, respectively. For $U = 12$, the $T_{MIT}$ exceeds the $T_N$
significantly. Additionally, we depict the temperature dependence of the specific heat $C_v$
for $U = 12$ in Fig.~\ref{fig01}(b) to demonstrate that the $T_N$ and the low temperature
($T \sim 0.2$) peak related to spin-fluctuations correspond well to each other. The moment
formation and the $T_{MIT}$ are related to the high temperature peak (around $T = 1$) in $C_v$ associated
with charge-fluctuations. In general, for the $U = 12$ case, an intervening
paramagnetic insulating region separates the PM-M and AF-I phases. On the other hand,
Fig.~\ref{fig01}(c) illustrates that for $U = 6$, the $T_N$ and the $T_{MIT}$ are comparable.
To analyze the relation between magnetic and transport properties further we plot the local
magnetic moment ($M$) as a function of temperature in Fig.~\ref{fig01}(d). As the temperature
decreases for $U = 8$, the magnetic moment $M$ increases and saturates at low temperatures.
It also displays a small peak near the magnetic transition, i.e., around $T_N$.
A very similar scenario is also observed for the $U = 12$ case as shown in the same figure. 
For $U = 6$, the $M$ remains more or less same as we approach $T_N$ from $T \sim 1$, but
increases below $T_N$. For smaller $U$ values, such as $U = 4$, the $M$ values even decrease
when the temperature decreases from $T \sim 1$ to $T_N$. The system remains PM-M until
$T_N$ for low $U$ values, possibly due to the delocalization of magnetic moments.

\begin{figure}[t]
\centering
\includegraphics[width=0.48\textwidth]{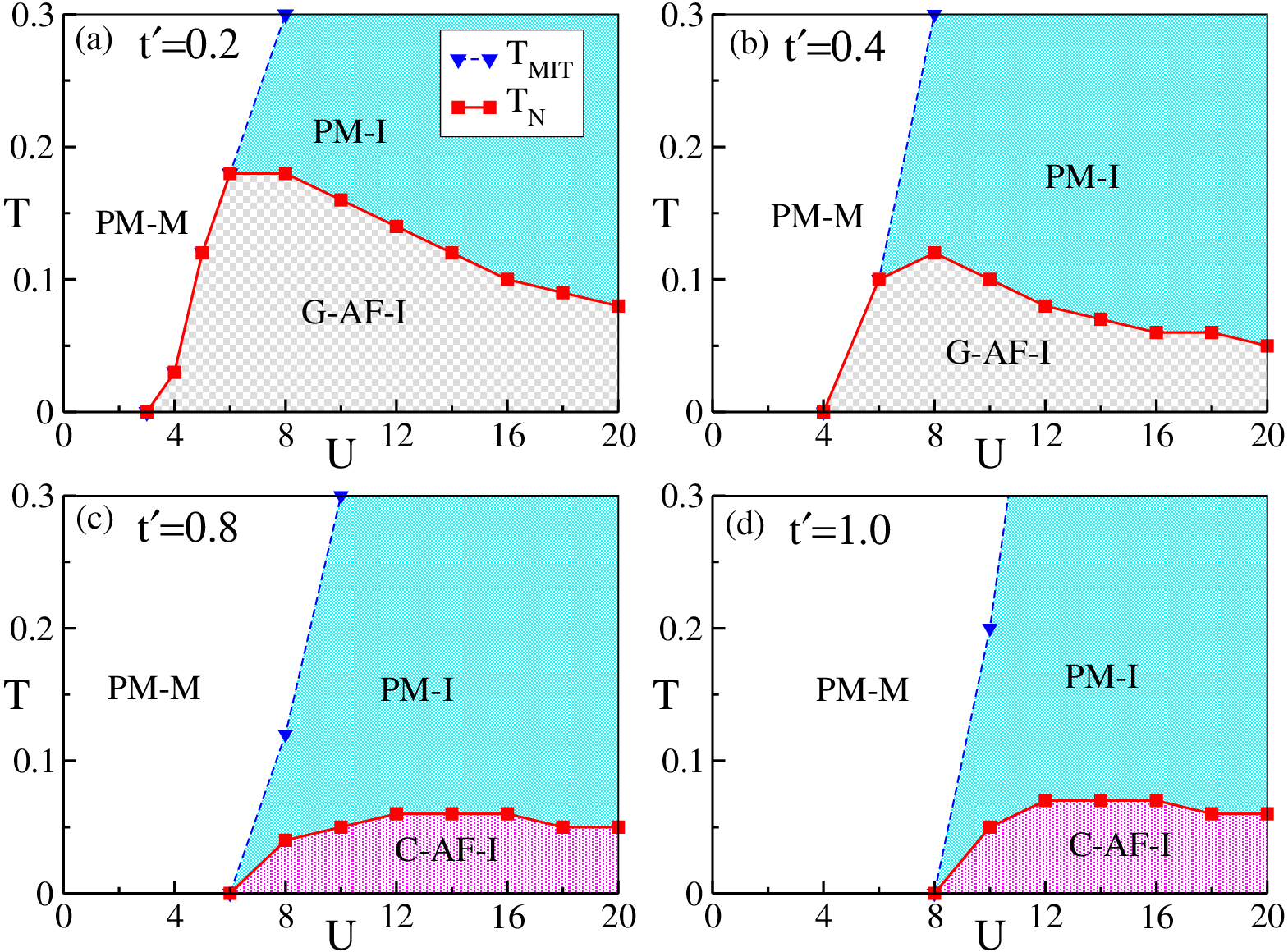}
\caption{The $U$-$T$ phase diagrams at finite $t'$ values:
(a) $t' = 0.2$, (b) $t' = 0.4$, (c) $t' = 0.8$, and (d) $t' = 1.0$. As $t'$ increases,
the long-range AF order in the low-$U$ regime decreases, favoring the paramagnetic
metallic (PM-M) state at low temperatures. At $t' = 0.2$ and $t' = 0.4$, the G-type AF
insulating phase vanishes away at lower $U$ values, but remains stable at moderate to
high $U$ values. For high $t'$ [$t' = 0.8$ and $t' = 1.0$], the G-type AF insulating
phase disappears completely, whereas a C-type AF insulating phase appears for
moderate and high $U$ values. Legends are same in panels (a)–(d).
}
\label{fig02}
\end{figure}

\section{$t' \ne$ 0 while $t'' = $ 0: Effect of the second-nearest-neighbor hopping on magnetotransport properties}

The addition of second-nearest-neighbor hopping ($t'$) introduces frustration by competing
with nearest-neighbor hopping $t$. For $U \leq 3$, the G-type AF insulating phase totally
disappears as shown in Fig.~\ref{fig02}(a). When $3 < U \leq 6$, the system changes from the
PM-M phase to the G-type AF insulating phase as the temperature decreases. For $U \geq 8$,
the PM-I phase appears between the PM-M and G-type AF phases, similar to the $t' = 0$ case.
It is also important to observe that, in contrast to the non-frustrated situation ($t' = 0$),
competitive interactions reduce the transition temperature of the G-type antiferromagnetic
phase for all $U$ values when $t' = 0.2$ is added, though only slightly at higher $U$
values. However, the Neel temperature ($T_N$) exhibits non-monotonic behavior, similar
to the $t' = 0$ situation. As $t'$ increases to 0.4, the system stays paramagnetic at the
lowest temperature even at $U = 4$ as shown in Fig.~\ref{fig02}(b). For $U > 4$, the G-type
AF phase is stable at low temperatures. However, the antiferromagnetic transition temperature
decreases dramatically for all $U$ values as compared to $t' = 0$ and $t' = 0.2$ cases.
The suppression of $T_N$ due to $t' = 0.4$ is more noticeable for intermediate $U$
values compared to higher $U$ values. These all findings highlight the role that competitive
interactions between $t$ and $t'$ play in suppressing the G-type AF insulating phase; that
is, a larger $t'$ results in a more noticeable suppression of the G-type AF phase.

As the $t'$ value is increased further, the effect of $t'$ turns into even more dramatic
and noticeable. At $t' = 0.8$, G-type AF order is entirely wiped out, resulting in the onset
of C-type antiferromagnetic order at higher $U$ values, as shown in Fig.~\ref{fig02}(c).
The C-type AF phase, which forms when $U > 6$, is insulating, similar to the G-type AF phase.
Thus, as the temperature rises for $t' = 0.8$, the system transits from the C-type AF
insulating phase to the PM-I phase at higher $U$ values. When $U \leq 6$, long-range magnetic
ordering disappears, resulting in the paramagnetic metallic phase, even at low temperatures.
The long-range magnetic order does not survive up to $U = 8$ at $t' = 1.0$, as shown in Fig.~\ref{fig02}(d).
The C-type AF insulating phase appears beyond that $U$ value.
So, in the nonperturbative regime, where the on-site Hubbard repulsive strenght U is comparable
to the electronic bandwidth of the system, introducing significant second-nearest-neighbor
hopping ($t'$) shifts the ground state preference from G-type to C-type AF.

\begin{figure}[t]
\centering
\includegraphics[width=0.48\textwidth]{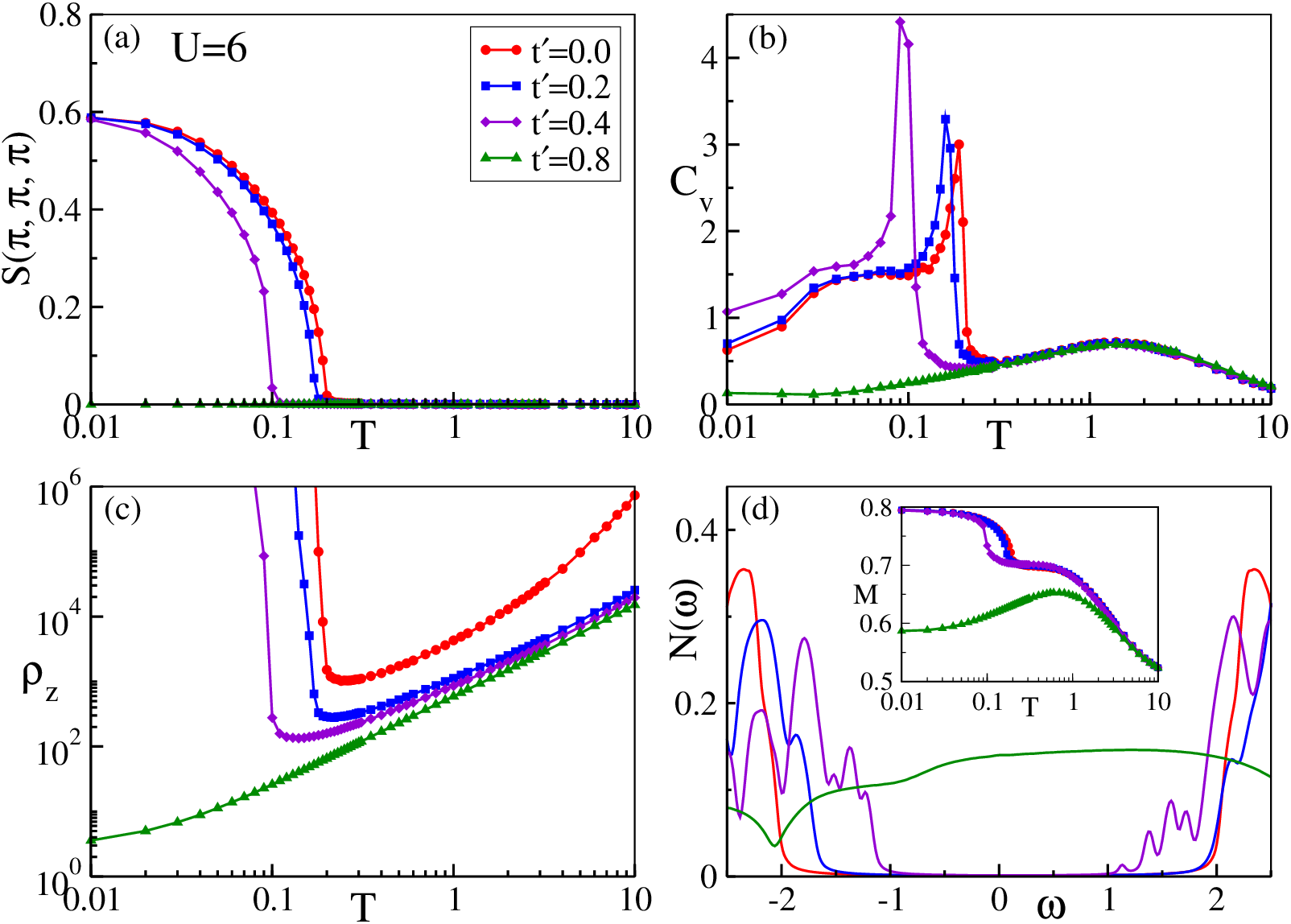}
\caption{
Physical quantities for $U = 6$ with various $t'$ values: (a) The evolution of
structure factor $S(\pi,\pi,\pi)$ with temperature for different $t'$ values. As
$t'$ increases, the $T_N$ decreases and eventually disappears for large $t'$.
(b) As $t'$ increases from 0 to 0.4, the low temperature peak in $C_v$ vs. $T$,
indicating the antiferromagnetic transition, shifts to lower temperatures. This
peak, however, vanishes at $t' = 0.8$, which is in line with the observation
that the magnetic ordering does not exist at this $t'$.
(c) Resistivity $\rho_z$ vs. T plots illustrate that when $t'$ increases, the
$T_{MIT}$ falls. At $t' = 0.8$, the system stays in a metallic state even at low
temperatures. For every fixed $t'$ value the $T_N$ and $T_{MIT}$ coincide at $U = 6$.
(d) The density of states (DOS) at $T = 0.01$ for $t' =0$, 0.2, and 0.4 displays
a gap around the Fermi level ($\omega=0$), confirming the insulating behavior at
low temperatures. As $t'$ increases, the gap decreases, and a finite DOS at Fermi
level for $t' = 0.8$ affirms the presence of metallic ground state. The evolution
of local moments $M$ with temperature is shown in the inset of (d). For
$t' = 0$, 0.2, and 0.4, $M$ steadily increases below $T_N$ as temperature decreases
further. On the other hand, when $t' = 0.8$, where $T_N$ disappears, $M$ continuously
decreases below $T \approx 1$, suggesting that the delocalized moment contributes
to the enhancement of its metallicity. Legends are same in panels (a)–(d).
}
\label{fig03}
\end{figure}

We then go into more detail about the magnetic and transport properties, which we utilized
to build the $U$-$T$ phase diagrams shown in Fig.~\ref{fig02}. We mainly concentrate on
$U = 6$ and $U = 12$ as in Fig.~\ref{fig01} for these analysis. First, we discuss
the magnetic and transport properties obtained for $U = 6$. The structure factor
calculations demonstrate that the $T_N$ for the G-type AF phase decreases as $t'$
increases, finally vanishing for large $t'$ [see Fig.~\ref{fig03}(a)]. In order to
determine how the specific heat ($C_v$) versus temperature for $U = 6$ corresponds
to the magnetic transition, it is also shown over a range of $t'$ values in
Fig.~\ref{fig03}(b). It is clear that the low-temperature peak around $T = 0.1$
corresponds to the respective $T_N$ for $t'= $0.0, 0.2, and 0.4. Our calculations
also show that the low temperature peak is absent at $t'= $ 0.8, which
is consistent with the fact that $T_N$ vanishes for this $t'$.

In addition, we investigated the temperature dependence of resistivity at various
$t'$ to determine the relationship between the long-range magnetic ordering and the
transport properties. The $T_N$ and the $T_{MIT}$ correspond well for $t'= $0.0, 0.2,
and 0.4 and vanish together at $t' = 0.8$, as shown in Fig.~\ref{fig03}(c). Thus,
for $t' = 0.8$, the system remains metallic down to low temperatures. In addition, we
calculate the density of states (DOS) at $T = 0.01$ to analyze the transport properties,
as illustrated in Fig.~\ref{fig03}(d). The DOS clearly shows a gap at the Fermi level
($\omega = 0$) for $t' = 0$, 0.2, and 0.4, indicating an insulating ground state.
However, the gap decreases with increasing the value of $t'$.
On the other hand, a finite DOS at the Fermi level for $t' = 0.8$ indicates a metallic
ground state. Furthermore, as the temperature decreases from $T \approx 1$, the local
magnetic moment ($M$) in the PM-M phase for $t' = 0.8$ decreases, indicating that a delocalized
moment enhances the metallicity, as illustrated in the inset of
Fig.~\ref{fig03}(d). However, for $t' = $ 0.2 and 0.4, the $M$ stays more or less
firm below $T = 1$ and increases below $T_N$, much like in the $t' =0$ case, giving
rise to insulating state at low temperatures.

\begin{figure}[t]
\centering
\includegraphics[width=0.48\textwidth]{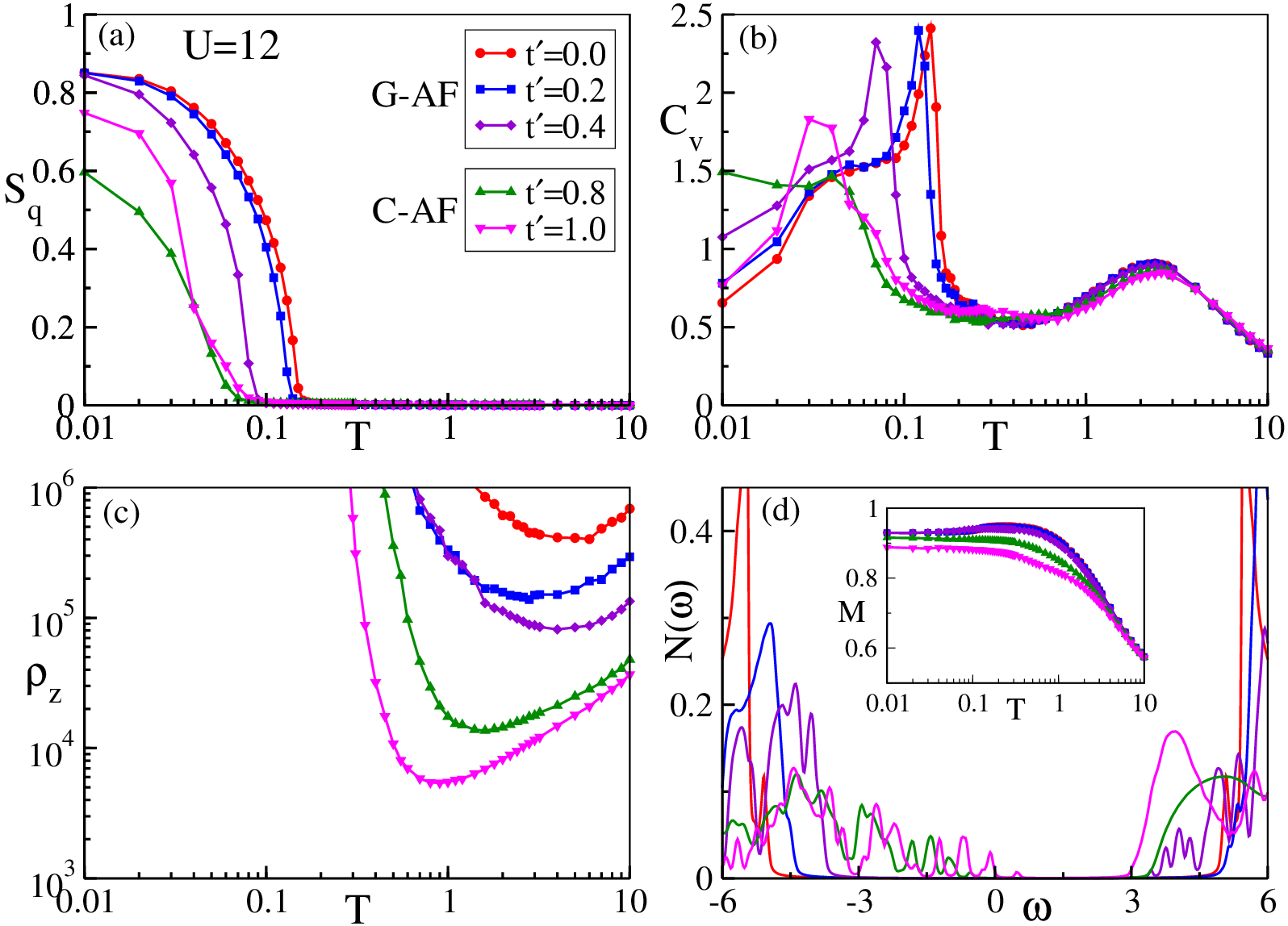}
\caption{ Physical quantities for $U = 12$ with different $t'$ values: (a) Structure factors
for G-type AF ordering are plotted at $t' = 0$, 0.2, and 0.4, while those for C-type
AF ordering are presented at $t' = 0.8$ and 1.0. When $t'$ (0, 0.2, and 0.4) is small,
the low temperature magnetic ordering is G-type AF, but $T_N$ decreases as $t'$ increases.
On the other hand, the low temperature magnetic order transforms into C-type AF for
stronger $t'$ (0.8 and 1) values. (b) The corresponding antiferromagnetic transitions
and the low-temperature peak in the $C_v$ vs. temperature curves are coherent.
(c) Temperature dependence of resistivity $\rho_z$ exhibits insulating behavior
at low temperatures for all $t'$ values. The $T_{MIT}$ decreases as $t'$ increases.
(d) Insulating behavior at low temperatures is highlighted by the density of states
(DOS) at $T = 0.01$, which displays a gap near the Fermi level ($\omega=0$).
This gap decreases as $t'$ increases. Satellite peaks emerge at the edge
of the gap in the C-type AF ordered phase, unlike the G-type AF ordered phase.
The temperature evolution of the local moment $M$ is illustrated in the inset of (d).
For every $t'$, $M$ rises with decreasing temperature and reaches saturation at
lower temperatures. The saturation value of $M$ at low temperatures decreases
gradually with increasing $t'$. Legends are same in panels (a)–(d).
}
\label{fig04}
\end{figure}

Now, we present a variety of physical signatures in Fig.~\ref{fig04} in order to investigate
the magnetotransport properties that appear at $U = 12$. The structure factor calculations
show that the $T_N$ for the G-type AF phase diminishes as $t'$ grows from 0.0 to 0.4,
eventually vanishing for large $t'$, as illustrated in Fig.~\ref{fig04}(a). The C-type AF
phase emerges at $t' = 0.8$ and $t' = 1.0$. In Fig.~\ref{fig04}(b), we show the
temperature-dependent specific heat ($C_v$) for $U = 12$ at different $t'$ values.
For the G-type phase in particular, the systematics of the low-temperature peak of $C_v$s
is in accordance with that of the $T_N$s demonstrated in Fig.~\ref{fig04}(a). Due to
competition between three different types of C-type ordering S($\pi, \pi, 0$),
S($0, \pi, \pi$), and S($\pi, 0, \pi$) near the phase transition, the low temperature
peak in the $C_v$ is either slightly off the mark or distorted for higher $t'$ values
where the C-type phase is stable.

The systematic of $T_{MIT}$ is illustrated in Fig.~\ref{fig04}(c). The $T_{MIT}$
decreases as $t'$ increases considerably. As previously mentioned, for $U = 12$, the
system stays in highly insulating state at low temperatures. It is also clear that, for
any $t'$, the $T_{MIT}$ is significantly higher than the corresponding $T_N$ values. We
show the DOS at $T = 0.01$ in Fig.~\ref{fig04}(d) to support our investigation of the
transport properties at low temperatures. Our calculations disclose a significant gap
at the Fermi level ($\omega = 0$), implying an insulating ground state for all $t'$
values. As temperature decreases from $T \approx 1$, the variation of local magnetic
moment ($M$) at finite $t'$, where G-type phase remains unaffected, resembles to that
of the $t' = 0$ scenario, as illustrated in Fig.~\ref{fig04}(d), suggest that the
insulating property of these systems is preserved. At low temperatures, the saturation
value of $M$ decreases as $t'$ increases beyond 0.4, which reflect that the competing
interactions between $t$ and $t'$ somewhat hinder the creation of local moments
even at $U = 12$. The drop in $T_{MIT}$, as illustrated in Fig.~\ref{fig04}(c),
is in good agreement with the decrease in the saturation value of $M$ at
low temperatures with an increase in $t'$ values.

\begin{figure}[t]
\centering
\includegraphics[width=0.48\textwidth]{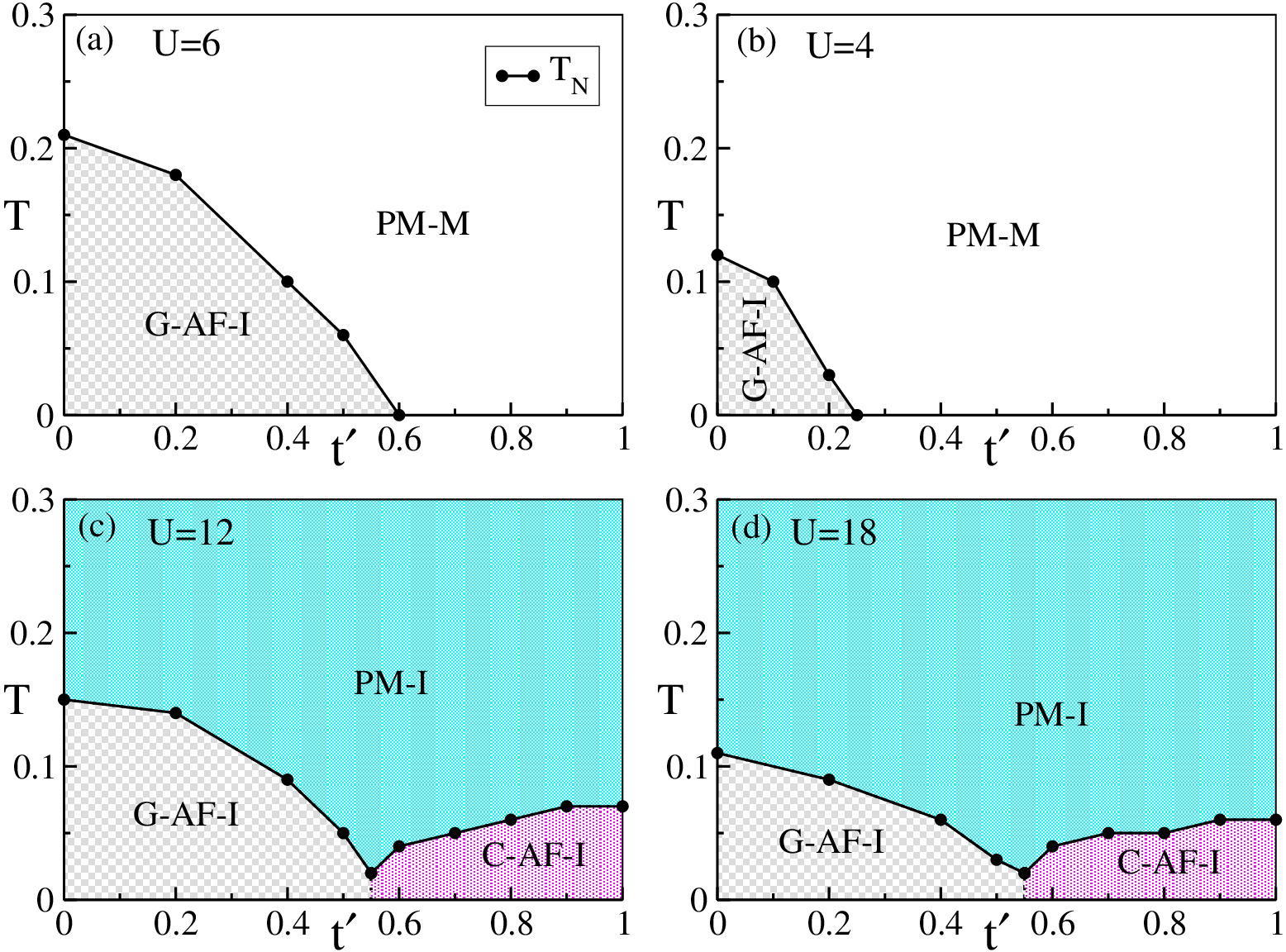}
\caption{
The $t'$-$T$ phase diagrams for different $U$ values:
(a) $U = 6$, (b) $U = 4$, (c) $U = 12$, and (d) $U = 18$. Compared to $U = 4$,
the region over which the G-type AF insulating phase remains stable,
is larger for $U = 6$. For smaller $U$ values, the $T_N$ and the $T_{MIT}$ coincide
even at finite $t'$, analogous to the $t' = 0$ scenario, as the system evolves
directly from a PM-M phase to a G-type AF insulating phase as the temperature decreases.
As $t'$ increases, the $T_N$ values fall until they vanish at $t' =$ 0.6 ($t' =$ 0.25)
at $U = 6$ ($U = 4$), indicating a low-temperature PM-M phase at higher $t'$ values.
Even at finite $t'$, the $T_{MIT}$ is much greater than $T_N$ for larger $U$ values
($U = 12$ and $U = 18$), due to the formation of a paramagnetic insulating (PM-I)
phase between the magnetically ordered insulating phase and the PM-M phase. As the
temperature decreases, the PM-I phase evolves to the G-type AF phase for smaller
$t'$ ($< 0.55$). Crucially, the PM-I phase transforms to a C-type AF insulating
phase when the temperature decreases for stronger $t'$ ($> 0.55$).
Legends are same in panels (a)–(d).
}
\label{fig05}
\end{figure}

In order to further illustrate the variation in magnetic phases at finite temperatures,
we present the $t'$-$T$ phase diagram for four $U$ values in Fig.~\ref{fig05}. The ground
state displays a long-range G-type AF insulating phase up to $t' = 0.6$ for $U = 6$, as
illustrated in Fig.~\ref{fig05}(a). The $T_N$ decreases as $t'$ increases and then
expectedly disappears. So, the competitive character of $t'$ prevents magnetic ordering
and makes it challenging to maintain long-range G-type AF order at low
temperatures, resulting in a decrease in $T_N$. In other words, beyond a certain $t'$,
competition between $t$ and $t'$ prohibits the system from achieving any long-range
magnetic order, retaining a paramagnetic metal (PM-M) as the ground state. Thus, the
system remains in a PM-M state due to conflicting
interactions that destabilize the antiferromagnetic order in this regime. At $U = 4$, the
G-type AF insulating state only lasts until $t' = 0.25$ [see Fig.~\ref{fig05}(b)].

The scenario is significantly different for higher $U$ values. For $U =$ 12 and 18,
we present the $t'$-$T$ phase diagram in Figs.~\ref{fig05}(c) and (d), respectively.
The ground state remains in the G-type AF insulating state for $t' < 0.55$. The C-type
AF insulating phase, in contrast to the PM-M phase in lower $U$ values, becomes
the ground state for $t' > 0.55$. The system shows the PM-I phase for all $t'$ values
just above the magnetic ordering temperature. Following the system's transition from
the G-type AF phase to the C-type AF phase at low temperatures, the $T_N$ of the C-type
AF phase initially increases and then remains more or less unchanged
with $t'$. These findings show that whereas weaker Hubbard interactions are more
susceptible to competing interactions that disrupt the magnetic ordering, strong
interactions ($U$) preserve a certain magnetic ordering (C-type AF) even at large $t'$.

\begin{figure}[t]
\centering
\includegraphics[width=0.48\textwidth]{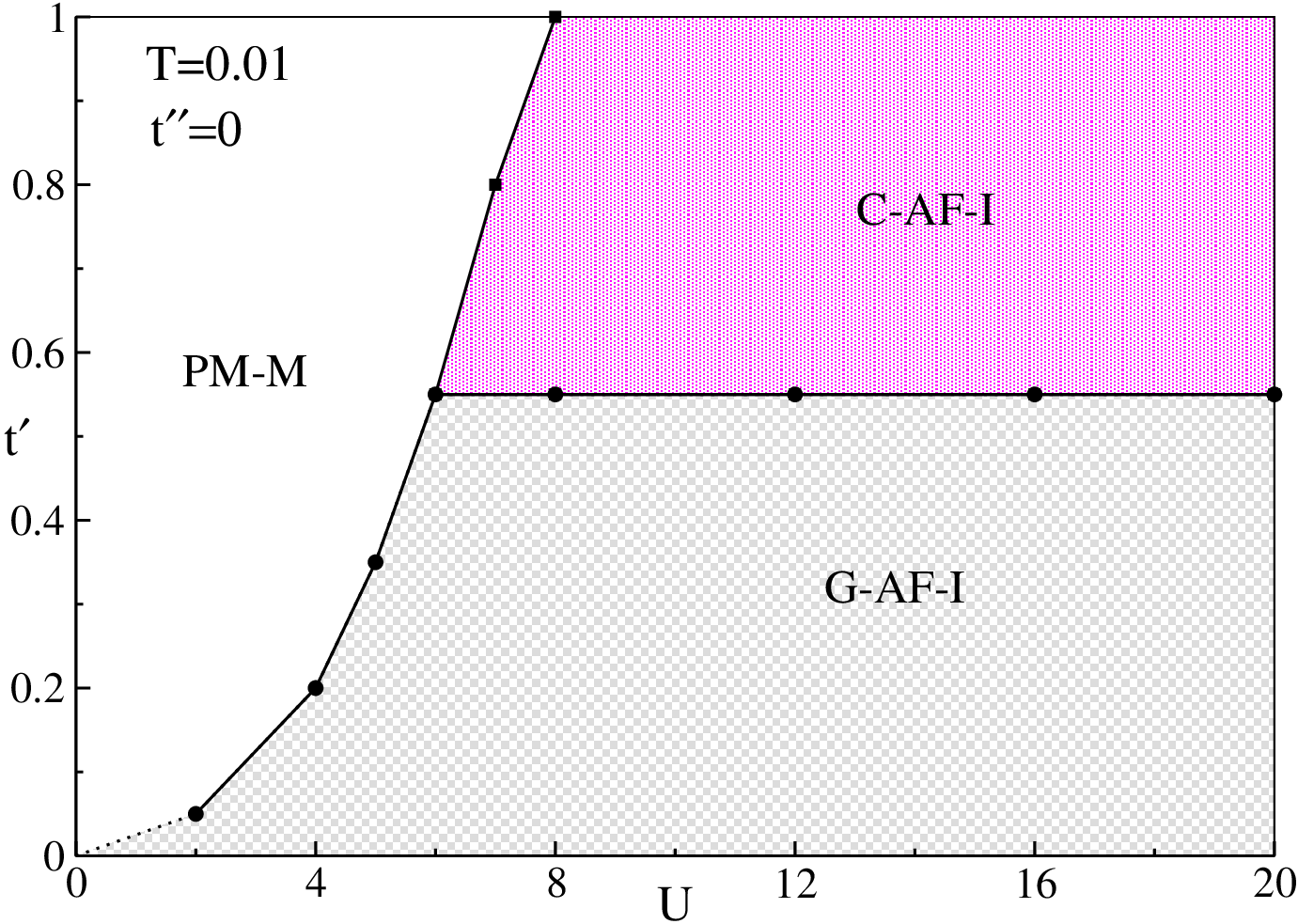}
\caption{The $U$-$t'$ phase diagram at $T = 0.01$: For $U \le 6$, the G-type AF phase
becomes destabilized and a PM-M phase becomes stabilized as $t'$ increases. For $U \ge 8$,
the G-type AF insulating phase transitions to the C-type AF insulating phase as $t'$
increases. The value of $t'$ where the G-type to C-type transition occurs, essentially
remains unchanged. For $U = 7$, the system transitions from G-type AF insulating to
PM-M phase through C-type AF insulating. $t''$ is set to zero in this plot.
}
\label{fig06}
\end{figure}

To completely demonstrate the impact of $t'$ across different ranges of $U$ values
[from weak ($U \ll BW$) to strong ($U \gg BW$)], we provide the ground state $U$-$t'$
phase diagram in Fig.~\ref{fig06} before we end this section. When $U$ is
small, the G-type AF insulating phase is the ground state due to the perseverance of perfect
nesting at $t'$ = 0. A small amount of $t'$ disturbs this nesting pattern, resulting in a PM
metallic phase. So, as $t'$ increases, the G-type AF insulating state changes to a PM-M state
at small and modest $U (\le 6)$ values. At intermediate and strong correlation limits
($U > 6$), the transition from G-type AF insulating to C-type AF insulating state
occurs approximately at $t' \approx 0.55$.

\begin{figure}[t]
\centering
\includegraphics[width=0.48\textwidth]{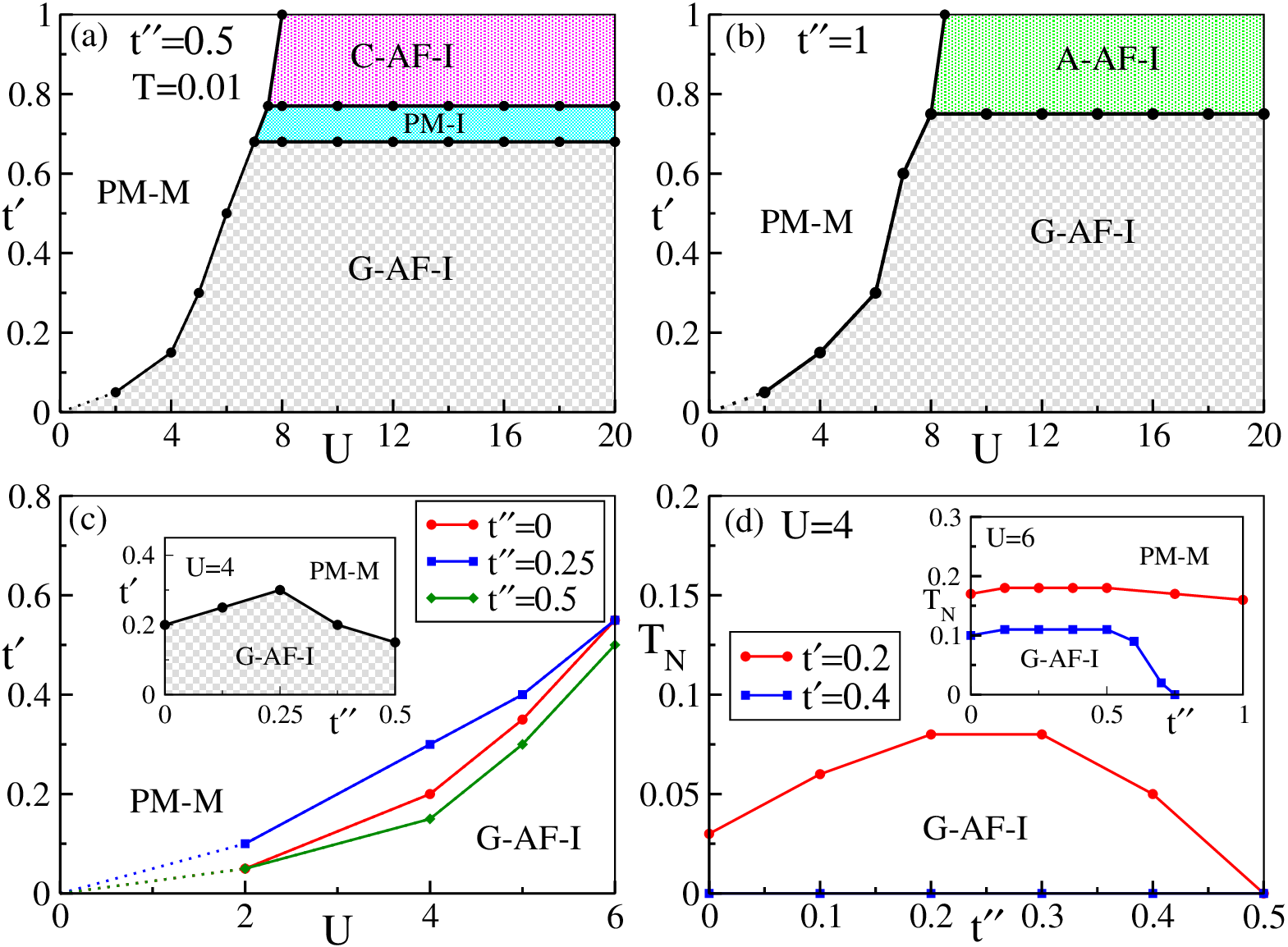}
\caption{The effect of $t''$: $U$-$t'$ phase diagrams for
(a) $t'' = 0.5$ and (b) $t'' = 1.0$ are displayed to compare to the case when
$t'' = 0$. When $t''=0.5$, a PM-I phase emerges within a small window of
$t'$ ($0.7 \le t' \le 0.75$) for $U > 7$, between the G-type and C-type AF
phases, as opposed to $t'' = 0$. Otherwise, the phase diagram appears similar
to $t'' = 0$. Surprisingly, at $t'' = 1$, the PM-I and C-type AF insulating
phases are entirely eliminated, whereas the A-type AF insulating phase emerges
for $U > 8$. (c) Comparison of the $U$-$t'$ phase diagram for $U \le 6$ for
small and modest $t''$ values ($t'' = 0$, 0.25, and 0.5). The G-type AF phase
is stable below each curve, whereas the PM-M phase appears above it. As $t''$
increases from 0 to 0.25, the G-type AF insulating phase expands, indicating
a better stability of the G-type AF order. However, the G-type AF phase
contracts to lower $t'$ values as $t''$ increases to 0.5, suggesting decreased
stability of the G-type AF order. The inset highlights the non-monotonic nature
of $t''$ for $U = 4$. The stability of the G-type AF order improves from $t'' = 0$
to $0.25$, enabling it to persist at higher $t'$ values, but the stability
decreases with further rise of $t''$, according to the $t''$-$t'$ phase diagram
in the inset. (d) The $t''$-$T_N$ plot for $t' = 0.2$ at $U = 4$ illustrates
nonmonotonic behavior of $T_N$ with $t''$. Magnetic order is not formed for any
$t''$ when $t' = 0.4$. Inset: The $t''$-$T_N$ plot for $U = 6$ indicates a
non-monotonic trend of $T_N$ at $t' = 0.4$. When $t' = 0.2$, the $T_N$ remains more
or less constant.
}
\label{fig07}
\end{figure}

\section{Combined effect of long-range hoppings ($t' \ne 0$ and $t'' \ne 0$) on magnetotransport properties}

In this section, we investigate various cases in which the third-nearest-neighbor
hopping ($t''$) is finite. We employ the same range of $U$ and $t'$ values that we
have used for $t''=0$ case in Fig.~\ref{fig06} to study the comprehensive ground
state $U$-$t'$ phase diagram for $t'' = 0.5$ [see Fig.~\ref{fig07}(a)]. In contrast
to the $t''=0$ scenario, a strongly correlated paramagnetic insulating (PM-I) phase
separates the G-type AF insulating and C-type AF insulating phases for intermediate
to large $U$ values at $t'' = 0.5$, emphasizing the complex interactions between
$t'$ and $t''$. On the other hand, the PM-I and C-type AF insulating phases totally
disappear at $t'' = 1.0$ [see Fig.~\ref{fig07}(b)], and the G-type AF insulating
phase switches directly to the A-type AF insulating phase as $t'$ increases. It is
also noticeable that the G-type AF insulating phase persists over a substantially
greater range of $t'$ than at $t'' = 0.5$, demonstrating the
stabilizing influence of higher $t''$ on the G-type AF phase at higher $U$ values.

To better understand the competition between $t$, $t'$, and $t''$ values, we
present the ground state $U$-$t'$ phase diagram for a smaller range of $U$ values
with different $t''$ values in Fig.~\ref{fig07}(c). In this diagram, the G-type AF phase is stable below
each of the curve, while the PM-M phase is obtained above it. For small
$U$ values, there is clear non-monotonic behavior in terms of stabilization of
the G-type AF phase. In order to visualize the situation, we depict
the $t''$-$t'$ phase diagram in the inset of Fig.~\ref{fig07}(c) for $U = 4$.
The G-type AF is not stable at $t' > 0.2$ while $t'' = 0$, but becomes stabilized
at $t'' = 0.25$.  So, a small but finite $t''$ reduces the negative effect
of $t' \ne 0$ on long-range magnetic ordering and helps in strengthening the G-type AF
insulating phase at slightly larger $t'$ values. However, further increasing
the value of $t''$ reduces this effect.

At $t'$ = 0.2, the $T_N$ also exhibits a non-monotonic trend as $t''$ increases
for $U = 4$ [see Fig.~\ref{fig07}(d)]. The G-type AF phase disappears at $t''$ = 0.5.
For $t' = 0.4$, the system remains in a PM state, regardless of the $t''$ values.
The inset of the same figure shows the variation of $T_N$ with $t''$ for
$U = 6$. The $T_N$ remains more or less stable with $t''$ upto $t''$ = 1.0 at
$t' = 0.2$ but vanishes beyond $t''$ = 0.75 for $t' = 0.4$.

\begin{figure}[t]
\centering
\includegraphics[width=0.48\textwidth]{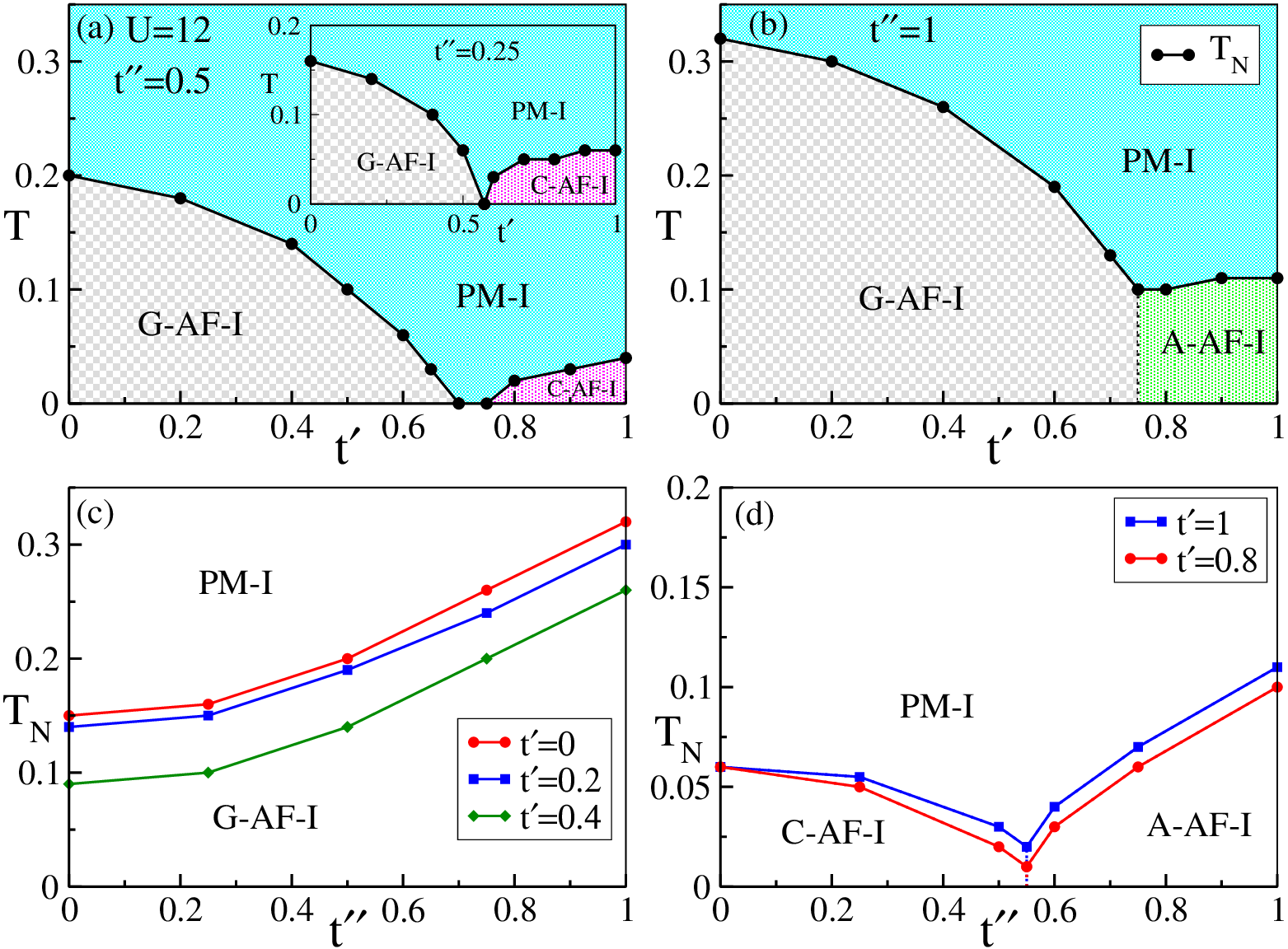}
\caption{The $t'$-$T$ phase diagrams at $U = 12$ are displayed for (a) $t'' = 0.5$ and
(b) $t'' = 1$.  We use the inset of (a) to illustrate the phase diagram for $t'' = 0.25$.
For all three values of $t''$, the $T_{MIT}$ is significantly higher than $T_N$.
resulting in the creation of the PM-I phase at temperatures above the magnetically
ordered insulating phase. At low to moderate $t'$, the PM-I phase transitions to the
G-type AF insulating phase, while at high $t'$, it transitions to either the C-type
AF insulating phase ($t'' = 0.25$ and $0.5$) or the A-type AF insulating phase ($t'' =$ 1.0),
depending on the value of $t''$. The low-temperature G-type AF phase ground state
remains stable for $t' < 0.55$, but a C-type AF insulating phase appears for
$t' > 0.55$ ($t' > 7$) for $t'' = $0.25 ($t'' = $ 0.5). For $t'' = 0.5$, in the
narrow $t'$ window ($0.7 \leq t' \leq 0.75$), the PM-I phase persists down to the
lowest accessible temperature, separating the low-$t'$ G-type AF insulating phase
from the high-$t'$ C-type insulating phase. At $t'' = 1$ the PM-I phase transitions
to the G-type AF insulating phase for $t' < 0.75$, while for $t' > 0.75$, it evolves
to the A-type AF insulating phase. (c) As $t''$ increases, the $T_N$ associated with
the G-type AF phase increases for any given $t'$. But, for a specific $t''$, the
$T_{N}$ decreases as $t'$ increases. (d) The $T_N$ decreases as $t''$ increases
in the C-type AF region for $t' = 0.8$ and 1. In contrast, the $T_N$ increases
with $t''$ inside the A-type AF ordered regime. Legends are same in panels (a) and (b).
}
\label{fig08}
\end{figure}

Next, we examine the $t'$-$T$ phase diagram for $t'' = $0.25, 0.5 and 1 in order to
better understand the impact of $t''$ for $U = 12$. For $t'' = 0.5$, as shown in
Fig.~\ref{fig08}(a), the $T_N$ decreases with increasing $t'$, similar to the $t'' = 0$
scenario. After $t' = 0.75$, the system shifts to a C-type AF insulating phase, and as
$t'$ increases, the $T_N$ for this phase is enhanced slightly. At low temperatures, the
transition from G-type to C-type AF phase occurs through the PM-I phase in our calculations.
Otherwise, the phase diagram at $t'' = 0.5$ is qualitatively similar to $t'' = 0$ scenario.
Additionally, we display the $t'$-$T$ phase diagram for $t'' = 0.25$ in the inset of
Fig.~\ref{fig08}(a). In this case, there is essentially no separation between
G-type AF and C-type AF phases near their junction similar to $t'' = 0$ case. As $t''$ increases
to 1, the G-type AF insulating phase persists until $t' = 0.75$, as shown in
Fig.~\ref{fig08}(b). Interestingly, for $t' > 0.75$, a long-range A-type AF insulating
phase appears instead of the C-type AF phase, as compared to $t'' = 0.5$ case.

When observing the systematics of the $T_N$ values of the G-type AF phase at small
and intermediate $t'$ values, it is clear that the $T_N$ values at $t''$ = 1 are
significantly higher than the values observed at $t''$ = 0 or 0.5. This increase
in $T_N$ with an increase in $t''$ values is more apparent from the $t''$-$T_N$ plot
for $t'$ = 0, 0.2, and 0.4 [shown in Fig.~\ref{fig08}(c)]. This highlights how
the $t''$ hopping strengthens the G-type AF phase by complementing the $t$ hopping
at $U = 12$. In addition, for $t'' = 1$, the Neel temperature of
the A-type AF phase is higher than the Neel temperature of the C-type phase that
appears in the $t'' = 0$ or $t'' = 0.5$ cases for a given $t'$ value.
In Fig.~\ref{fig08}(d), we depict the $t''$-$T_N$ phase diagram for $t'$ = 0.8 and 1.0 in order
to demonstrate the systematics of transition temperatures as we traverse from the
C-type AF to the A-type AF phase. First, the $T_N$ of the C-type AF phase decreases
as $t''$ grows and then, beyond $t'' = 0.55$, the A-type AF phase appears, and the $T_N$ of
A-type AF phase increases with increase in $t''$. 

\begin{figure}[t]
\centering
\includegraphics[width=0.48\textwidth]{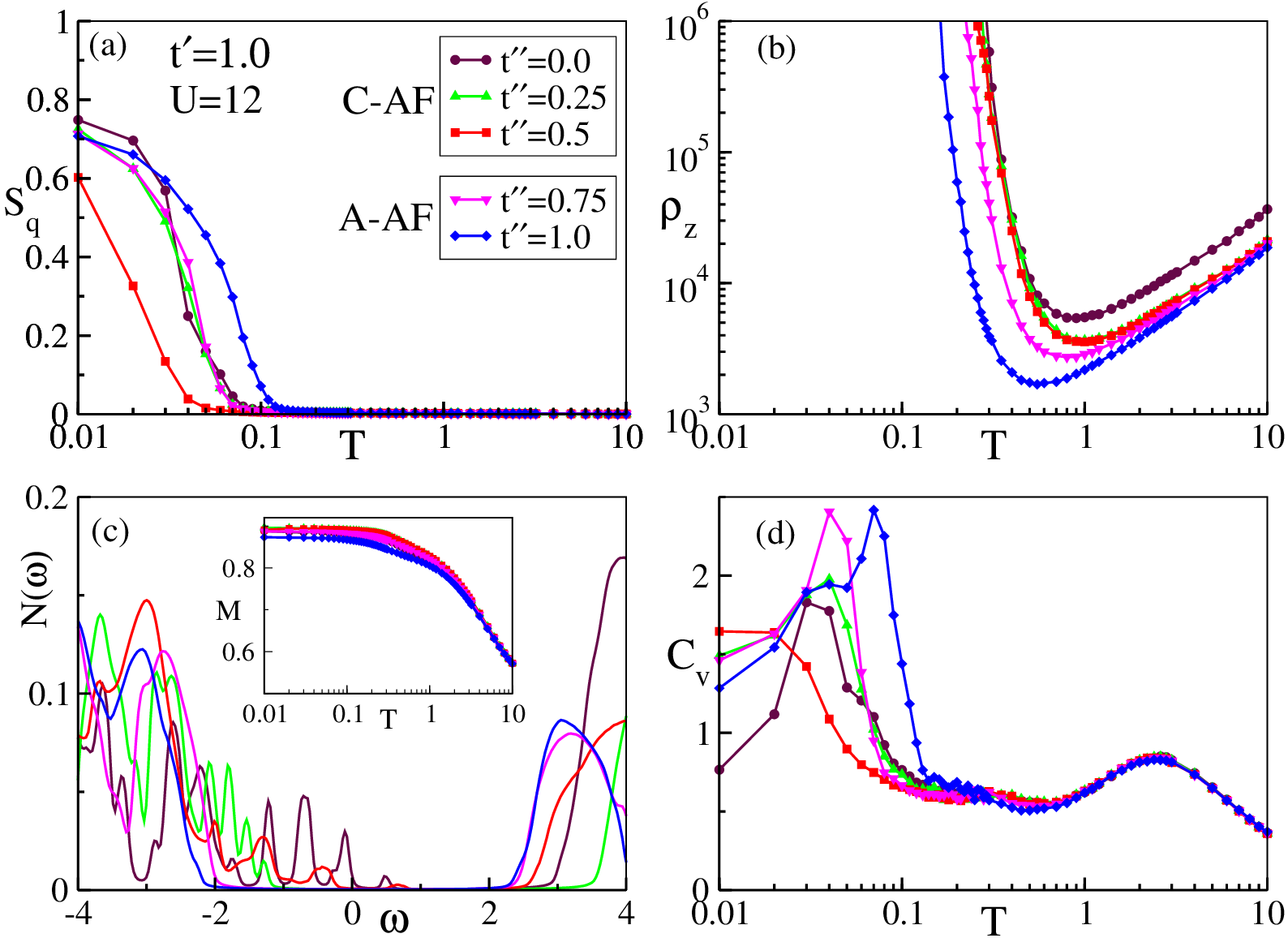}
\caption{Physical properties for $U = 12$ at $t' = 1.0$ with different $t''$
values: (a) Variation of the structure factor [$S(\pi,\pi,0)/S(\pi,0,\pi)/S(0,\pi,\pi)$]
for C-type AF order with temperature at $t'' = 0$, 0.25, and 0.5. The $T_N$ for C-type
AF phase decreases as $t''$ increases from 0 to 0.5. The structural factor for A-type
AF order [$S(\pi,0,0)/S(0,0,\pi)/S(0,\pi,0)$] vs T demonstrates that when $t''$ increases
from 0.75 to 1, the $T_N$ value increases. (b) Resistivity $\rho_z$ versus T is shown
for various $t''$. In the C-type AF ordered region, the $T_{MIT}$ slightly increases
or remains same as $t''$ increases from 0 to 0.5, whereas in the A-type ordered region, the $T_{MIT}$
decreases as $t''$ increases. However, the $T_{MIT}$ is significantly larger than
the $T_N$ for all $t''$ values. (c) At $T = 0.01$, the DOS shows a gap at the Fermi
level ($\omega=0$), confirming the insulating characteristic of all $t''$ values at
low temperatures. In the C-type AF ordered region, increasing $t''$ do not affect the
low temperature saturation value of local moments $M$, while increasing $t''$ in
the A-type AF ordered region decreases it. (d) The location of the low-temperature
peak in specific heat ($C_v$) varies in accordance with the $T_N$ shown in (a). 
Although the peak for the C-type AF phase becomes weaker. For more details please see the text.
Legends are same in panels (a)–(d).
}
\label{fig09}
\end{figure}

\begin{figure}[t]
\centering
\includegraphics[width=0.48\textwidth]{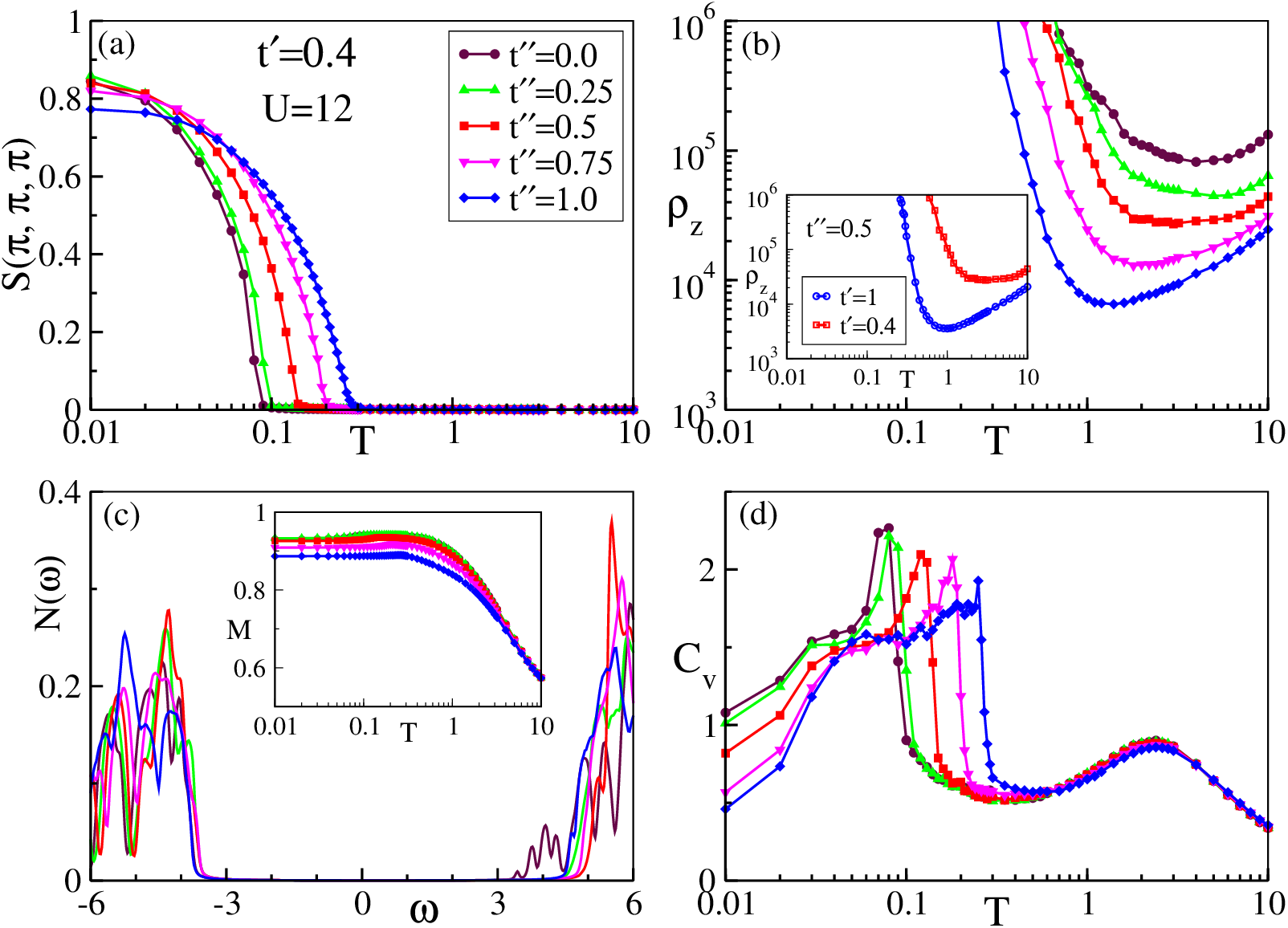}
\caption{Physical properties for $U = 12$ at $t' = 0.4$ for various $t''$ values:
(a) The temperature evolution of the structure factor $S(\pi,\pi,\pi)$ corresponding
to G-type AF ordering is plotted for $t''= 0$, 0.25, 0.5, 0.75, and 1. The $T_N$
increases with increasing $t''$. (b) Resistivity $\rho_z$ vs. temperature plots
demonstrate that when $t''$ increases, the $T_{MIT}$ decreases. Furthermore, the
$T_{MIT}$ is much greater than $T_N$ for all $t''$ values. In addition, $T_{MIT}$
values are significantly bigger for the $t' = 0.4$ case than for the $t' = 1$
scenario for any $t''$ value. Inset: $\rho_z$ vs. T are plotted for $t' = 0.4$ and
$t' = 1$ cases using $t''$ = 0.5 to compare the $T_{MIT}$ values. (c) For all $t''$,
the DOS at $T = 0.01$ shows insulating behavior at low temperatures by displaying a gap
around the Fermi level ($\omega=0$). The temperature variation in local moments $M$
is shown in the inset. At low temperatures, the saturation value of $M$ decreases
as $t''$ increases. (d) As $t''$ increases, the low-temperature peak in $C_v$
changes to higher temperatures, in line with the G-type AF transition temperature
illustrated in (a). Legends are same in panels (a)–(d).
}
\label{fig10}
\end{figure}

Next, using $U = 12$, we demonstrate more details of the magnetic and transport
properties with different $t''$ values for $t' = 1$ in Fig.~\ref{fig09}. The structure
factor calculations [see Fig.~\ref{fig09}(a)] show that the $T_N$ for the C-type AF
phase decreases as $t''$ increases, while the $T_N$ for the A-type AF phase increases
as $t''$ rises from 0.75 to 1. Interestingly, as shown in Fig.~\ref{fig09}(b), the
metal-insulator transition temperature $T_{MIT}$ increases marginally or remains same
as $t''$ increases within the C-type AF ordered region, but it decreases as $t''$ grows
inside the A-type AF phase. However, $T_{MIT}$ is always substantially larger
than the corresponding $T_N$ value. To further examine the transport properties,
we compute the density of states (DOS) at $T = 0.01$, as shown in Fig.~\ref{fig09}(c).
The DOS shows a gap at the Fermi level ($\omega = 0$) for all $t''$, indicating an
insulating ground state. It is crucial to notice that the satellite peaks observed
in the gap when $t' = 1$ was introduced [see Fig.~\ref{fig04}(c)] are gradually
disappeared as one increases $t''$. Additionally, inset of Fig.~\ref{fig09}(c)
shows how the local magnetic moment ($M$) essentially saturates at low temperatures
below $T \approx T_N$, indicating the insulating nature of the system. The
observation that $T_{MIT}$ decreases with $t''$ in the A-type AF ordered region
is in good agreement with the saturation value of $M$ at low temperatures,
which decreases with increasing $t''$. On the other hand, as $t''$ increases
within the C-type AF ordered region, the saturation value of $M$ remains more
or less same. We additionally exhibit the specific heat ($C_v$) with temperature
for different $t''$ values in Fig.~\ref{fig09}(d) to ascertain whether or not
the low temperature peak in the specific heat and the $T_N$ relate with one another.
In fact, the trend of the temperature at which the low-temperature peak of $C_v$
is observed for all $t''$ is roughly connected to the $T_N$ systematics in both
A-type and C-type AFs. The low-temperature peak of $C_v$ broadens around switching
points of the magnetic phases, as previously mentioned.

We also present the magnetotransport properties for $t' = 0.4$ for comparison in
Fig.~\ref{fig10}. The structure factor calculations show that the $T_N$ for the
G-type AF phase increases with increasing $t''$, as shown in Fig.~\ref{fig10}(a).
This is in contrast to the $t' = 1$ scenario, in which the system initially stays
in the C-type AF phase (i.e., for small and intermediate $t''$ values) before
transitioning to the A-type AF phase. In Fig.~\ref{fig10}(b), it is clear that
when $t''$ increases, the $T_{MIT}$ decreases. Similar to the $t' = 1$ scenario,
$T_{MIT}$s are significantly larger than $T_N$s. In addition, we calculate the
DOS at $T = 0.01$ in Fig.~\ref{fig10}(c). For every $t''$, the DOS shows a
distinct large gap at the Fermi level ($\omega = 0$), in contrast to the $t' = 1$
scenario. It is also noticeable that the $T_{MIT}$ values are significantly
bigger for the $t' = 0.4$ case than for the $t' = 1$ scenario for any $t''$
value [see inset of Fig.~\ref{fig10}(c) for $t''$ = 0.5],
which is consistent with the large gap. Additionally, the $M$ vs. temperature
curve in the inset of Fig.~\ref{fig10}(c) reflects the systematics of $T_{MIT}$.
The saturation magnetization at low temperatures is significantly higher than
that of the $t' = 1$ cases shown in the inset of Fig.~\ref{fig09}(c).
Furthermore, for $t' = 0.4$, the saturation magnetization at low temperatures
decreases with increasing $t''$, which is consistent with a decrease in $T_{MIT}$ shown
in Fig.~\ref{fig10}(b). We additionally demonstrate the temperature-dependent
specific heat ($C_v$) for different $t''$ values in Fig.~\ref{fig10}(d) for
completeness. The $T_N$ values correspond to the temperature where low-temperature
peak of $C_v$ occur for all $t''$ values. Increasing $t''$ shifts the
low-temperature peak of the $C_v$ to higher temperatures, analogous to the
enhancement of $T_N$ as shown in Fig.~\ref{fig10}(a).

\begin{figure}[t]
\centering
\includegraphics[width=0.48\textwidth]{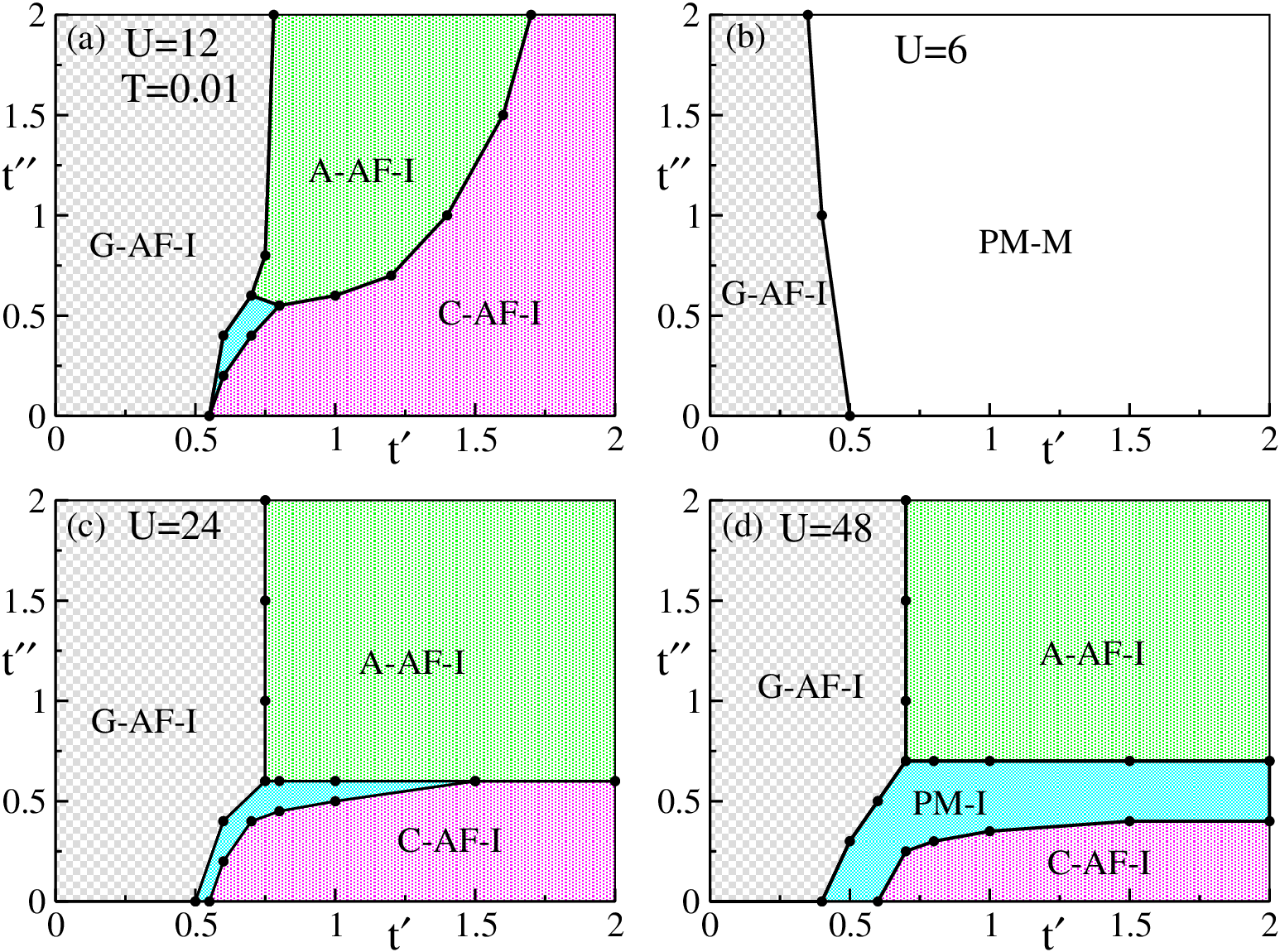}
\caption{$t'$-$t''$ phase diagrams for different $U$ values at $T = 0.01$:
(a) $U = 12$, (b) $U = 6$, (c) $U = 24$, and (d) $U = 48$.
When $t'' = 0$, the system directly transitions from the G-type AF insulating phase
to a C-type AF insulating phase with increasing $t'$ for $U = 12$. As $t''$ increases,
a PM-I phase separates the G-type and C-type AF insulating phases. When $t'' > 0.6$,
the PM-I phase disappears and the G-type AF insulating phase changes directly to the
A-type AF insulating phase. As $t'$ value increases further, the system eventually
transitions from the A-type AF insulating phase to the C-type AF insulating phase.
However, the C-type AF-I phase that occurs when large $t'$ and $t''$ values are
coupled at $U = 12$ is not observed for $U = 24$ or $U = 48$. As $U$ grows from
12 to 24 to 48, the correlated PM-I phase at the intersection of three AF phases
expands, primarily suppressing the C-type AF-I phase. For $U = 6$, the phase diagram
in (b) demonstrates that only one kind of long-range magnetic phase persists.
Beyond a reasonable $t' \sim$ 0.5, long-range magnetic ordering breaks down for
all $t''$ values, stabilizing the PM-M phase.}
\label{fig11}
\end{figure}

Now we demonstrate the $t'$-$t''$ phase diagrams at low temperature ($T =0.01$) for
different on-site repulsion strengths $U$ in Fig.~\ref{fig11} to comprehend the
fine aspects of magnetotransport properties. For $U = $12, the G-type AF phase is stable
for small and moderate $t'$ values, as shown in Fig.~\ref{fig11}(a). If one observes
closely, the G-type AF's disappearance shifts toward larger $t'$ values as $t''$
increases for $t'' \lessapprox$ 0.75. This happens as a result of $t''$ maintaining
the stability of the G-type AF phase by supporting $t$, the nearest-neighbor hopping.
On the other hand, even in this small and intermediate $t''$ ($\lessapprox$ 0.55)
range, the C-type AF becomes stabilized at higher $t'$ values. Significantly, when
$t'$ increases, the system evolves from the G-type AF phase to the C-type AF phase
via a paramagnetic insulating (PM-I) phase in intermediate $t''$ regime. At higher
$t''$ values, i.e., $t''\gtrapprox 0.55$ the system switches directly to the A-type
AF phase from the G-type AF phase as $t'$ increases. At very large $t'$ values,
however, the system eventually converts to the C-type AF phase. Thus, our whole
phase diagram illustrates how multiple magnetic phases in a half-filled system
compete with one another due to a variety of conflicting interactions for $U = $12.
Importantly, the PM-I phase that appears near the junction of three different types of
magnetic insulating phases persists into the strong-coupling limit. The lack of
magnetic order down to $T = 0.01$ characterizes this region. So, the occurrence can be
attributed to a quantum-disordered regime of the local moments. Interestingly, the
PM-I phase emerges in intermediate $t'$ and $t''$ values, disrupting the balance
between the AF phases at the phase boundaries. Therefore, the additional fluctuation
processes provided by $t''$ successfully destroy long-range order and preserve the
insulating character, even though the competition between $t$ and $t'$ alone is
insufficient to accomplish this. However, its presence is restricted to a small region.
The formation of the PM-I state can be traced to quantum fluctuations
in strongly correlated materials~\cite{Laubach}.

We also illustrate the $t'$-$t''$ phase diagram for $U = 6$ in Fig.~\ref{fig11}(b).
The phase diagram shows only one long-range magnetic phase:
the G-type AF phase. Long-range magnetic order breakdown occurs beyond a moderate
$t'$ value (e.g., $t' \sim 0.5$ for $t'' = 0$), stabilizing the low-$U$ PM-M phase.
Increasing $t''$ does not dramatically affect the boundary between the G-type AF
insulating and PM-M ground states, which remains mostly unaltered.

The competition between the
various phases shown for $U = $12 is maintained when the $U$ value is increased
to 24 as shown in Fig.~\ref{fig11}(c). The C-type AF phase, which was stable for
large $t'$ and $t''$ values at $U = 12$, is no longer stable at $U = 24$. However,
the A-type AF phase is extended to that regime. In this case, a PM-I phase also emerges
between the junction of G-, C-, and A-type AF phases. This PM-I phase separates the
C-type and A-type AF phases as the $t''$ value is increased for a fixed $t'$ value
near $t' = 1$. Overall, the area covered by the PM-I phase for $U = 24$ is larger
than for $U = 12$.

In contrast to $U = 12$, it is important to note that the PM-I phase has
apparently stabilized for a range of $t'$, albeit small, even if $t'' = 0$. The
range across which the PM-I phase is stabilized at $t'' = 0$ gets broader when
the $U$ value is raised to 48 [see Fig.~\ref{fig11}(d)]. Increasing $U$ values
also enriches the PM-I region at $U = 48$ while reducing the C-type AF
phase. Otherwise, the $t'$-$t''$ phase diagram looks similar to that of $U$ = 24
[compare Fig.~\ref{fig11}(d) and (c)].

\begin{figure}[t]
\centering
\includegraphics[width=0.48\textwidth]{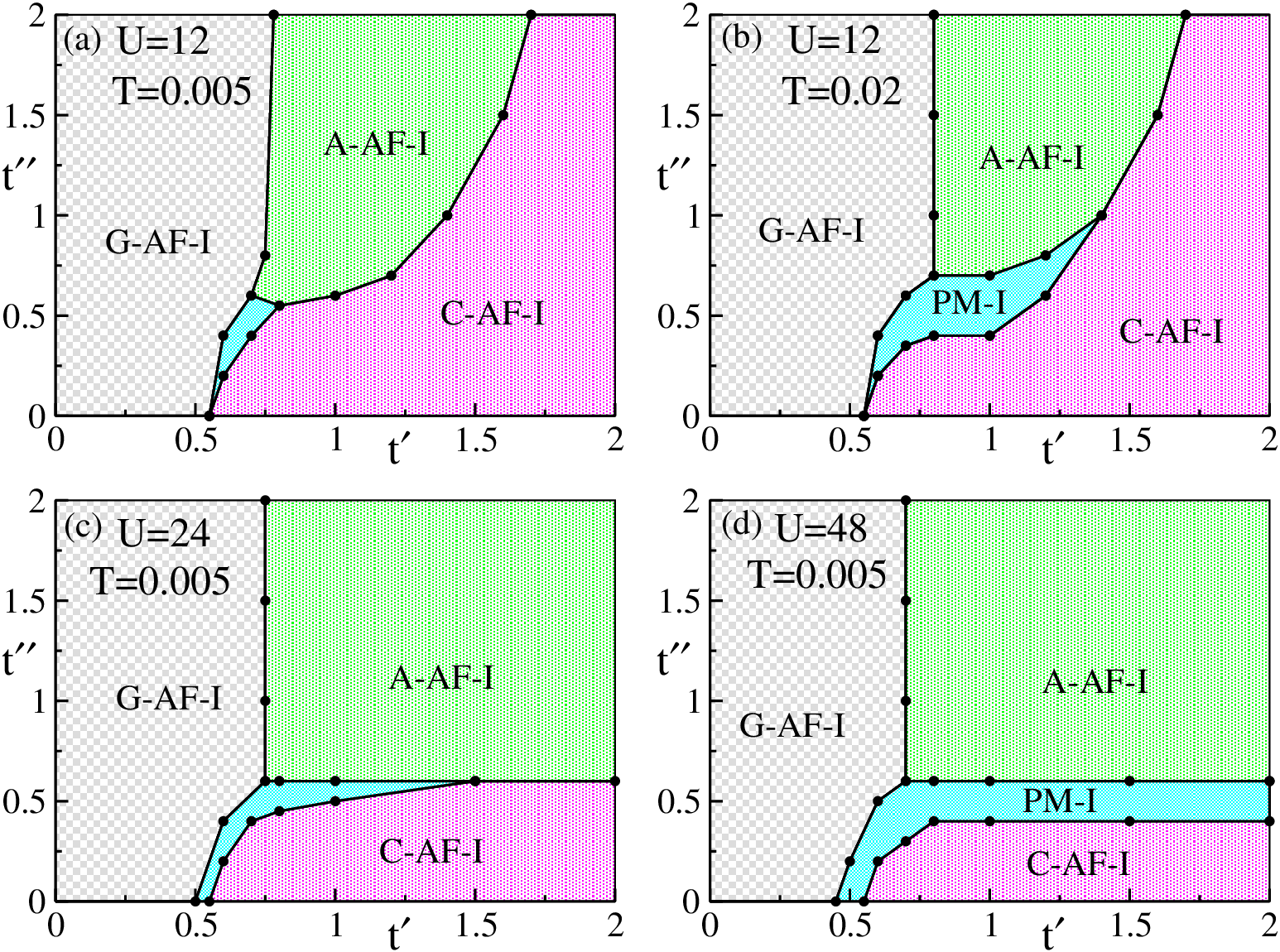}
\caption{$t'$-$t''$ phase diagrams for various combinations of $U$ and $T$ values:
(a) For $U = 12$ at $T = 0.005$; (b) for $U = 12$ at $T = 0.02$;
(c) for $U = 24$ at $T = 0.005$; (d) for $U = 48$ at $T = 0.005$. The phase diagram for
$U = 12$ at $T = 0.005$ is very similar to the phase diagram at $T = 0.01$ illustrated in
Fig.~\ref{fig11}(a). In contrast to $T = 0.01$ and $0.005$, the PM-I region is amplified
at $T = 0.02$. The phase diagram at $T = 0.005$ remains unchanged from the one at $T = 0.01$
in Fig.~\ref{fig11}(c), when $U = 24$. On the other hand, the PM-I phase region for $U = 48$
at $T = 0.005$ is somewhat smaller than the PM-I phase at $T = 0.01$ in Fig. \ref{fig11}(c).
}
\label{fig12}
\end{figure}

Overall, as shown in Fig.~\ref{fig11}, our calculations reveal that strong electron
correlations and long-range magnetic interactions can combine to produce a variety
of magnetic phases at low temperatures. Interestingly, low temperatures, long-range
interactions, and quantum processes can all combine to produce novel paramagnetic
insulating phases that go against expectations. Is the PM-I phase that we observed
at $T = 0.01$ stable at even lower temperatures? In order to answer this, we plot
the $t'$-$t''$ magnetic phase diagram for $U = 12$ at $T = 0.005$ and $T = 0.02$ in
Fig.~\ref{fig12}(a) and Fig.~\ref{fig12}(b), respectively. The PM-I region captured
at $T = 0.01$ is evidently unaltered at $T = 0.005$, but it enlarges for $T = 0.02$.
This demonstrates that when the temperature decreases, the PM-I phase remains remarkably
stable. The stability of the PM-I phase at lower temperatures is also demonstrated for
$U = 24$ and $U = 48$ in the Figs.~\ref{fig12}(c) and (d), respectively.

When a PM-I phase occurs at low temperatures in presence of long-range interactions
it raises the possibility of a spin liquid state. In other words, the
combination of long-range interactions and paramagnetic insulation may produce
ideal conditions for the formation of spin liquid states in strongly correlated systems. 
Overall, a PM-I phase with long-range interactions may support a spin liquid state, 
but this is not guaranteed. Also, stabilizing spin-liquid states in three dimensions is 
more challenging than in two dimensions due to weakening of quantum fluctuations as 
dimensionality increases. So, the identification of a potential host for spin-liquid 
phase in a three-dimensional Hubbard model with long-range interactions is also interesting.

Similar non-magnetic insulating regions in the phase diagram have also been
found in earlier studies~\cite{Tocchio, Tsirlin, Isaev, Majumdar, Jiang, Hu, Gong}
employing various methods such as the variational cluster approach~\cite{Laubach}.
These results further suggest the possibility that this system might
host three-dimensional quantum spin liquid phases. Given the long-range
interactions in a strongly correlated electron system, our findings are
consistent with previous studies that suggested possibility of a spin-liquid
phase. However, we agree that advanced quantum many-body techniques, such as 
Density Matrix Renormalization Group (DMRG) or Quantum Monte Carlo (QMC), 
can provide a more accurate description of the ground state and low-energy excitations 
in the paramagnetic insulating region, leading to a better understanding of their nature.

\section{Conclusion}

Using the s-MC approach, we examined the effects of
long-range hoppings ($t'$ and $t''$) beyond the nearest-neighbor hopping $t$ on
the electronic and magnetic properties of the half-filled Hubbard model on a cubic
lattice. We performed an extensive study of the processes in which long-range
hopping and the degree of electron-electron interaction combine to drive
different magnetic phase transitions at low temperatures. The finite-temperature
evolution of the G-, A-, and C-type AF  phases as well as the underlying
metal-insulator transitions were investigated thoroughly. In non-perturbative limit,
where electron-electron interaction $U \sim$ bandwidth, the $T_{MIT}$ is much
larger than the $T_N$, resulting in a PM-I phase at temperatures higher than
$T_N$. Interestingly, a novel PM-I phase is created by the combined influence
of long-range interaction and quantum processes at very low temperatures. This
PM-I phase emerges in intermediate $t'$ and $t''$ values and lies at the
intersection of three distinct AF insulating phases. Our calculations
clearly demonstrate that the additional fluctuating processes, introduced by $t''$, provides
actually aid in triggering the collapse of long range order while maintaining the
insulating nature of the system at low temperatures, even though the competition
between $t$ and $t'$ is generally insufficient to accomplish this at $U \sim$
bandwidth. This PM-I phase percolate to $t''$ = 0 for $U >>$ bandwidth limit.  
Does this PM-I phase host a 3D spin-liquid phase? Validating the spin-liquid
phase requires detailed theoretical computations of spin-spin interactions and
excitations, which are beyond the scope of this paper. We leave this as a topic
for future studies. Overall, our findings are significant because they shed light
on the complex behavior of magnetic systems in the presence of long-range
interactions in correlated electron systems, which has implications for
exploration and designing of novel materials with exotic electronic and
magnetic properties. The finding of a potential candidate for spin-liquid phase
in a three-dimensional Hubbard model with long-range interactions is
particularly noteworthy. The theoretical insights from this work may
provide useful direction for experiments aimed at creating novel materials
with particular magnetic properties or those that could potentially
host a three-dimensional spin-liquid state.


\section*{Acknowledgment}
We acknowledge use of the Meghnad2019 computer cluster at SINP.


\appendix
\section{Effective spin-fermion Hamiltonian} \label{derivation_Heff}
To initiate the semiclassical Monte-Carlo (s-MC) technique, we begin by dissociating
the interaction term in Eq.~\ref{hamiltonian} through the conventional
Hubbard–Stratonovich (HS) transformation, introducing auxiliary fields. This involves
expressing the interaction term at each lattice site in a decouple manner as demonstrated below,
\begin{align} \label{dh_1}
n_{i\uparrow}n_{i\downarrow} = \frac{1}{4} n_i^2 - S_{iz}^2 = \frac{1}{4} n_i^2 - \left( \mathbf{S}_i.{\hat{\Omega}}_{i} \right)^2.
\end{align}
In this context, the spin operator at site $i$ is defined as
$\mathbf{S}_i = \frac{1}{2} \sum_{\alpha\beta} c_{i\alpha}^\dag \sigma_{\alpha\beta} c_{i\beta}$, where $\hbar=1$. 
Here, $\sigma = (\sigma_x,\sigma_y,\sigma_z)$ represents the Pauli matrices.
Additionally, ${\hat{\Omega}}_{i}$ denotes an arbitrary unit vector at site i.
We leverage the rotational invariance of
$S_{iz}^2$, thus, $\left( \mathbf{S}_i.{\hat{\Omega}}_{i} \right)^2 = S_{ix}^2 = S_{iy}^2 = S_{iz}^2$.

The partition function for the Hamiltonian is given by $Z = \mathrm{Tr}[\exp(-\beta H)]$,
where $H$ corresponds to the Hamiltonian stated in Eq.~\ref{hamiltonian}. The trace is
performed over the occupation number basis, encompassing all particle numbers and site
occupations. Here, $\beta = 1/T$ represents the inverse temperature, with the Boltzmann
constant $k_B$ set to $1$. The interval $[0,\beta]$ is segmented into $M$ equidistant
slices, each separated by an interval $\Delta \tau$, where $\beta = M\Delta \tau$. 
Utilizing Suzuki-Trotter decomposition, as $\Delta \tau$ tends towards zero for very large value of $M$, we
express $ \exp \left[-\beta (H_0 + H_I)\right] $ approximately as
$\left[ \exp(-\Delta \tau H_0) \exp(-\Delta \tau H_I) \right]^M $ up to the first order
in $\Delta \tau$.

Utilizing Eq.~\ref{dh_1}, for a given time slice `$l$', the interacting part of the partition function 
can be demonstrated to be proportional to
\begin{align}
&\int d \phi_i(l) d \Delta_i(l) d^2\Omega_i(l) \exp \left( - \Delta \tau \left\{ \sum_i \left [ \frac{{\phi_i(l)}^2}{U} + i\phi_i(l) n_i \right. \right. \right. \nonumber \\
&\left. \left. \left. + \frac{{\Delta_i(l)}^2}{U} -2\Delta_i(l) {\hat{\Omega}}_i(l) \cdot \mathbf{S}_i  \right ] \right\} \right),
\end{align}
where, $\phi_i(l)$ represents the auxiliary field associated with charge density, $\Delta_i(l)$
stands for the auxiliary field linked with spin density, and $\hat{\Omega}_i(l)$ is an arbitrary unit vector. 
Furthermore, we define a new vector auxiliary field $\mathbf{m}_i(l)=\Delta_i(l) {\hat{\Omega}}_i(l)$.
$\phi_i(l)$ couples to the local charge density operator, while $\mathbf{m}_i(l)$ couples
to the local spin operator. Consequently, the full partition function is proportional to
\begin{align}
&\mathrm{Tr} \prod_{l=M}^1  \int d\phi_i(l) d^3\mathbf{m}_i(l) \exp \left( - \Delta \tau \left\{ H_{0} + \sum_i \left[ \frac{{\phi_i(l)}^2}{U} \right. \right. \right. \nonumber \\
& \left. \left. \left. + i\phi_i(l) n_i
+ \frac{{\mathbf{m}_i(l)}^2}{U} -2\mathbf{m}_i(l) \cdot \mathbf{S}_i  \right] \right\} \right),
\end{align}
where, the integrals are taken over $\{ \phi_i(l), \mathbf{m}_i(l) \}$. 
The product order for the slices ranging from $l=M$ to $l=1$ implies time-ordered
products over time slices, with earlier times positioned to the right. It is important
to note that at this stage, the partition function is exact, showcasing SU(2) symmetry,
and the $\{ \phi_i(l), \mathbf{m}_i(l) \}$ fields fluctuate in both space and imaginary
time. Now we proceed by making the following approximations: (i) retaining solely the
spatial dependence of the auxiliary field variables by dropping the $\tau$ dependence
of auxiliary field, and (ii) employing a saddle point value for the auxiliary fields,
$i\phi_i = \frac{U}{2} \langle n_i \rangle$. Then, with the redefinition
$\mathbf{m}_i \to \frac{U}{2} \mathbf{m}_i$, the above approximation facilitates the
derivation of the effective spin-fermion type Hamiltonian $H_{\text{eff}}$ as stated
in Eq.~\ref{h_eff}, where fermions couple to classical auxiliary fields $\{ \mathbf{m}_i(l) \}$.

\section{Observable definitions} \label{obs}

To determine the ground state magnetic ordering, we calculate the quantum correlations
$S(\textbf{q})$, defined as,
\begin{align}
S(\textbf{q}) = \frac{1}{(L^3)^2} \sum_{kl} \left\langle \textbf{s}_k \cdot \textbf{s}_l \right\rangle \exp\left[-i\textbf{q} \cdot (\textbf{r}_k - \textbf{r}_l) \right].
\end{align}
Here, the angular brackets imply quantum and thermal averaging at all individual
sites. The indices $\{k,l\}$ sum over all $L^3$ sites. The wave vector $\textbf{q}$
characterizes various magnetic orderings:  
$\textbf{q} = (\pi, \pi, \pi)$ denotes G-type AF order, 
$\textbf{q} = (0, \pi, \pi), (\pi, 0, \pi), (\pi, \pi, 0)$ signifies C-type AF order, and 
$\textbf{q} = (0, 0, \pi), (\pi, 0, 0), (\pi, 0, 0)$ stands for A-type AF order. 

The average local moment, defined as the square of the quantum magnetization, is
expressed as,
\begin{align}
M = \left\langle (n_{\uparrow} - n_{\downarrow})^2 \right\rangle = \left\langle n \right\rangle - 2 \left\langle n_{\uparrow} n_{\downarrow} \right\rangle .
\end{align}
Here $\left\langle n \right\rangle = \left\langle n_{\uparrow} + n_{\downarrow} \right\rangle $ represents the average occupation number of the system. 
In the limit $U \to \infty$ at any finite temperature, the double occupation 
$\left\langle n_{\uparrow} n_{\downarrow} \right\rangle \to 0$. 
At half-filling, where $\left\langle n \right\rangle = 1$, this leads to $M \to 1$.
For the uncorrelated case ($U=0$) or at very high temperatures ($T \to \infty$) for
finite $U$, the double occupation satisfies
$\left\langle n_{\uparrow} n_{\downarrow} \right\rangle \to  \left\langle n_{\uparrow} \right\rangle \left\langle n_{\downarrow} \right\rangle$.
At half filling which implies
$\left\langle n_{\uparrow} \right\rangle = \left\langle n_{\downarrow} \right\rangle =0.5 \Rightarrow M = 0.5$.
Thus, for any real system at half-filling, the local moment $M$ lies between
0.5 and 1.

The specific heat of the system is calculated by differentiating the total energy with
respect to temperature (by employing the central difference formula for the derivative),
\begin{align}
C_v(U,T)=\frac{dE(U,T)}{dT}.
\end{align}
To study transport properties, we calculate the density of states (DOS) and resistivity. 
The DOS is defined as 
\begin{align}
  N(\omega)
&= \sum_{i=1}^{2L^3} \delta(\omega - \epsilon_i),
\end{align}
where $\epsilon_i$ are the single-particle eigenvalues of the effective
Hamiltonian~\ref{h_eff}. We use a Lorentzian representation of the delta function with a
broadening of $\sim BW/(2L^3)$ to evaluate the DOS, where $BW$ represents
non-interacting bandwidth and $L^3$ is the total number of lattice sites.

The resistivity is determined by inverting the d.c. limit of the optical conductivity, which
is obtained using the Kubo-Greenwood formula~\cite{Chakraborty, Mahan, Kumar2}.
The optical conductivity is defined as,
\begin{align}
\sigma(\omega) = \frac{A}{L^3} \sum_{\alpha, \beta} \left( n_\alpha -n_\beta \right) \frac{{\left\lvert f_{\alpha \beta} \right\rvert}^2}{\epsilon_\beta - \epsilon_\alpha} \delta \left[ \omega - \left( \epsilon_\beta - \epsilon_\alpha \right) \right],
\end{align}
with $A = \frac{\pi e^2}{\hbar a_0}$ ($a_0$ is the lattice parameter). The $f_{\alpha \beta}=\left\langle \psi_\alpha \left\vert \hat{j}_z \right\vert \psi_\beta \right\rangle$ are the matrix elements for the current operator $\hat{j}_z=ia_0 \sum_{i,\sigma} [t ( c_{i,\sigma}^\dag c_{i+a_0\hat{z},\sigma} -h.c.)+t'( c_{i,\sigma}^\dag c_{i+a_0\hat{z}+a_{0}\hat{x},\sigma} -h.c.)+t'( c_{i,\sigma}^\dag c_{i+a_0\hat{z}+a_{0}\hat{y},\sigma} -h.c.)+t''( c_{i,\sigma}^\dag c_{i+a_0\hat{x}+a_0\hat{y}+a_0\hat{z},\sigma} -h.c.)]$.
Here, $\epsilon_\alpha$ and $\left\vert \psi_\alpha \right\rangle$ are the eigenvalues and associated eigenvectors of the effective Hamiltonian~\ref{h_eff}, respectively, and $n_\alpha = f(\mu - \epsilon_\alpha)$ is the Fermi function. 
The average d.c. conductivity $\sigma_{dc}$ is calculated by integrating over a small
frequency window $\Delta\omega$,
\begin{align}
\sigma_{dc} = (\Delta\omega)^{-1} \int_0^{\Delta\omega} \sigma(\omega) \, d\omega.
\end{align}
$\Delta\omega$ is set four to five times larger than the system's mean finite
size gap (average eigen value separation), which is calculated as the ratio of
bandwidth $BW$ to total number of eigen values.

\end{document}